\documentclass[twocolumn,dvipsnames,prd,nofootinbib,floatfix,superscriptaddress]{revtex4}
\usepackage{xcolor}
\usepackage{amsmath,amssymb}
\usepackage{appendix}
\usepackage{subfigure}
\usepackage{natbib}
\usepackage{todonotes}
\usepackage{graphicx}
\usepackage{hyperref}
\usepackage{footnote}

\hypersetup{
    colorlinks=true,
    linkcolor=black,
    filecolor=magenta,      
    urlcolor=blue,
    pdftitle={Overleaf Example},
    pdfpagemode=FullScreen,
    }

\usepackage{dcolumn}
\usepackage{bm}
\usepackage{orcidlink}
\usepackage{hyperref}
\usepackage{xcolor}
\usepackage{soul}
\sethlcolor{yellow}

\begin{document}
\title{Hint towards inconsistency between BAO and Supernovae Dataset: The Evidence of Redshift Evolving Dark Energy from DESI DR2 is Absent}

\author{Samsuzzaman Afroz \orcidlink{0009-0004-4459-2981}}\email{samsuzzaman.afroz@tifr.res.in}
\author{Suvodip Mukherjee \orcidlink{0000-0002-3373-5236}}\email{suvodip@tifr.res.in}
\affiliation{Department of Astronomy and Astrophysics, Tata Institute of Fundamental Research, Mumbai 400005, India}

\begin{abstract}
The combination of independent cosmological datasets is a route towards precision and accurate inference of the cosmological parameters if these observations are not contaminated by systematic effects. However, the presence of unknown systematics present in differrent datasets can lead to a biased inference of the cosmological parameters. In this work, we test the consistency of the two independent tracers of the low-redshift cosmic expansion, namely the supernovae dataset from Pantheon$+$ and the BAO dataset from DESI DR2 using the distance duality relation which is a cornerstone relation in cosmology under the framework of General Relativity. We find that these datasets violate the distance duality relation and show a signature of redshift evolution, hinting toward unaccounted physical effects or observational artifacts. Coincidentally this effect mimics a redshift evolving dark energy scenario when supernovae dataset and DESI datasets are combined without accounting for this inconsistency. Accounting for this effect in the likelihood refutes the previous claim of  evidence of non-cosmological constant as dark energy model from DESI DR2, and shows a result consistent with cosmological constant with $w_0= -0.92\pm 0.08$ and $w_a= -0.49^{+0.33}_{-0.36}$. This is further supported by an increased Bayes factor at the value of the dark energy equation-of-state (EoS) for cosmological constant  ($w_0 = -1$, $w_a = 0$) when the distance duality inconsistency is accounted for. This indicates that the current conclusion from DESI DR2 in combination with Pantheon$+$ is likely due to the combination of two inconsistent datasets resulting in precise but inaccurate inference of cosmological parameters. In the future, tests of this kind for the consistency between different cosmological datasets will be essential for robust inference of cosmological parameters and for deciphering unaccounted physical effects or observational artifacts from supernovae and BAO datasets.
\end{abstract}

\maketitle

\section{Introduction}

The combination of independent cosmological datasets has long been recognized as a pathway to both precise and accurate determination of fundamental cosmological parameters \citep{2013PhR...530...87W,Lahav:2024npe,Staicova:2025huq,LuisBernal:2018drn,Piras:2024dml,Steinhardt:2025znn}. Independent measurements such as those from supernovae type Ia (SNIa), baryon acoustic oscillations (BAO), and the cosmic microwave background (CMB) have historically converged on a consistent model of the universe with dark matter and dark energy, apart from the disagreement in the value of the current expansion rate of the Universe (known as the Hubble constant) inferred from low redshift and high redshift probes \citep{Abdalla:2022yfr,Carr:2021lcj,DES:2018paw,DESI:2025zgx,Planck:2018vyg,Lemos:2023rdh,ACT:2025fju,DiValentino:2025sru}.

The advent of DESI marks the beginning of a new era in high-precision BAO observations, as it measures the large-scale clustering of galaxies and quasars across an extensive redshift range \citep{DESI:2025zgx,DES:2018gui,DESI:2016fyo}. The DESI DR2 results are combined with the low redshift luminosity distance measurement from SNIa from the Pantheon+ \citep{Brout:2022vxf}, Union \citep{Rubin:2023ovl}, and DESY5 \citep{DES:2024jxu} datasets to improve the precision on the inference of the low redshift expansion history of the Universe and hence obtaining tighter constraints on the dark energy equation-of-state (EoS) using the Chevallier-Polarski-Linder (CPL) parameterization \citep{Chevallier:2000qy,Linder:2002et,dePutter:2008wt}.  This data in combination with other cosmological probes denoted a strong evidence toward a evolving dark energy, and shows evidence towards ruling out cosmological constant with a statistical significance of 2.8 to 4.2 after  including BAO measurements with CMB and supernovae \cite{DESI:2025zgx}. There results delivers compressed distance observables that, after rigorous internal consistency checks of the line of sight and angular BAO measurements \citep{DESI:2024uvr}. Though such parametrization is a simple step towards exploring the dark energy evolution, its connection with the theoretical models are often questioned \citep{Shlivko:2024llw, PhysRevD.108.103519,Afroz:2024lou,Colgain:2024xqj,Colgain:2024ksa,Colgain:2024mtg,Colgain:2025nzf} and maybe more physics-driven model are essential to discover the dark energy EoS. However, a more crucial point to scrutinize to gauge the validity of this inference is the internal consistency of the different cosmological datasets used for deriving the dark energy EoS. 

In this study, we use the consistency test based on the cosmic distance duality relation (CDDR), which is valid under the General Theory of Relativity for any expansion history of the Universe \citep{Holanda:2010vb,Liao:2015uzb,Keil:2025ysb}. The CDDR connects the luminosity distance and the angular diameter distance, as expressed in Equation~\ref{eq:DistDuality}, as a function of cosmological redshift. Any two independent datasets of the cosmological distances say luminosity distance from supernovae and angular diameter distance from BAO, with the source redshifts inferred spectroscopically, should satisfy the CDDR. It is important to note that this relation holds for any dark energy EoS, according to the Etherington’s reciprocity theorem \citep{2007GReGr..39.1055E}. However, if there are unknown physical or unaccounted systematic effects in the observational data, a violation of this relation is then expected. As a result, CDDR can provide a physics-based consistency test between the datasets and any signature of disagreement of this consistency relation can hint towards breakdown of at least one of the assumptions made it the analysis. 

\begin{figure}[ht]
    \centering
    \includegraphics[width= 8.5cm, height =5.0cm]{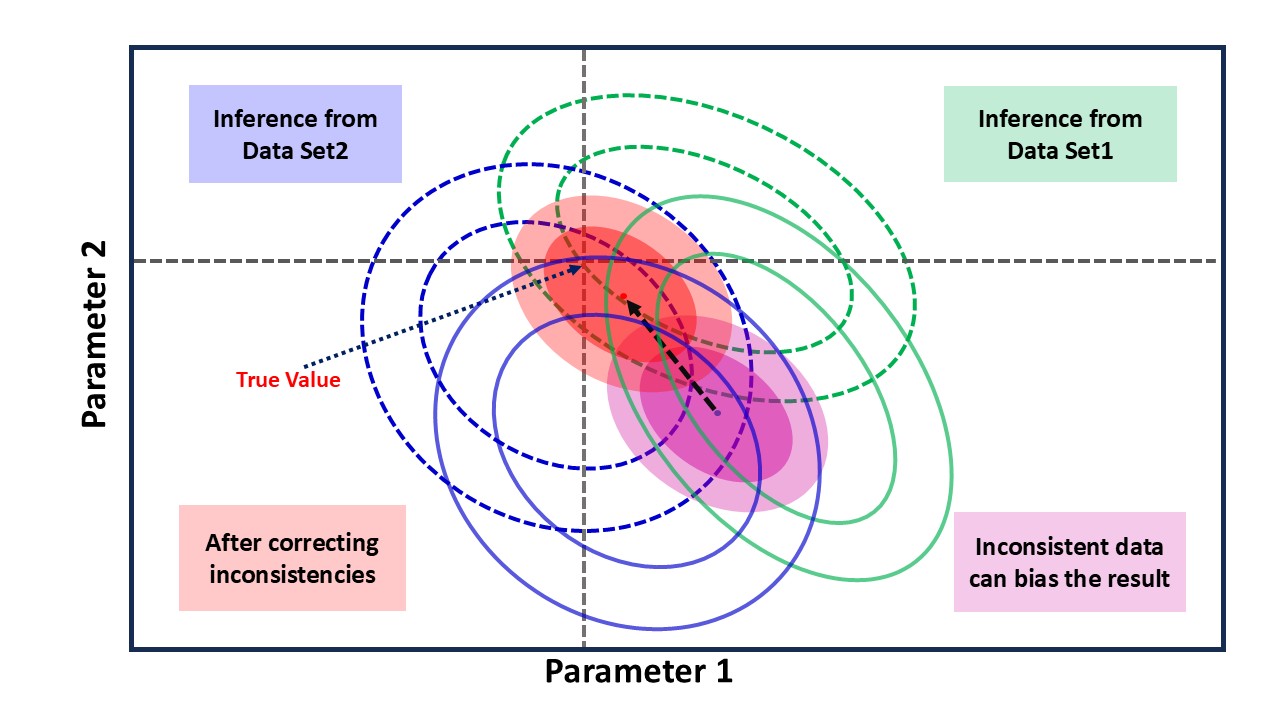}
    \caption{Illustration of how parameter inference from two datasets which are biased can can cause precise but an inaccurate inference. The dashed ellipses (green and blue) represent unbiased, independent constraints derived separately from Data Set 1 and Data Set 2, respectively. The solid ellipses (green and blue) indicate the same parameters measured in the presence of uncorrected systematic inconsistencies, causing shifts away from the true parameter values. The magenta filled ellipse shows the biased combined constraint resulting from naively merging these inconsistent measurements. After identifying and correcting the inconsistencies, the corrected combined inference is represented by the red filled ellipse, which realigns closely with the true parameter values. This schematic highlights the necessity of checking and correcting for inter-dataset inconsistencies prior to performing joint cosmological analyses. The boxes in different color explain different contours shown in the figure.}
    \label{fig:MPlot}
\end{figure}

The use of the CDDR offers several distinct advantages when testing for consistency between cosmological distance measurements derived from different observational probes:
\begin{itemize}
    \item \textbf{Model-independence:} The validity of CDDR is guaranteed for any cosmological expansion history within metric theories of gravity, allowing it to serve as a theory-agnostic consistency check.
    \item \textbf{Calibration sensitivity:} Since CDDR relates luminosity and angular diameter distances at the same redshift, any deviation signals residual calibration issues or redshift-dependent systematics in either dataset.
    \item \textbf{Systematics diagnosis:} A violation of the relation directly indicates the presence of unknown physical effects or observational systematics, without requiring assumptions about cosmological models.
    \item \textbf{Degeneracy mitigation:} Identifying and correcting inconsistencies via CDDR helps disentangle cosmological inference from dataset-specific systematics, leading to more robust parameter constraints.
\end{itemize}
These features make CDDR a powerful tool for joint analyses involving multiple distance measurements, such as those from supernovae and baryon acoustic oscillation datasets. We demonstrate this aspect on how the combining incorrect posteriors on the cosmological parameters can lead to a biased inference by a schematic diagram in Figure \ref{fig:MPlot}.

In this work, we apply this consistency test on the latest DESI DR2 release along with the SNIa dataset (Pantheon+) and found a \textit{redshift-dependent breakdown of the CDDR relation} by these two datasets, which indicates towards any unknown physical or systematic effect that is present in the datasets. Moreover, we find that the observed discrepancy can mimic a redshift evolving dark energy model and can bias the inferred value of the dark energy EoS. We further explore the inference of the dark energy EoS parameters along with the Hubble constant and matter density, and find that observed discrepancy in the datasets is strongly degenerate with the cosmological parameters. It is important to reiterate that though the breakdown of CDDR seen in the datasets is a cosmological model-independent statement, its presence can bias the cosmological results as luminosity distance and angular diameter distance will drive towards a values away from the true cosmological parameters. 

\begin{figure}[ht]
    \centering
    \includegraphics[width= 8.5cm, height =5.5cm]{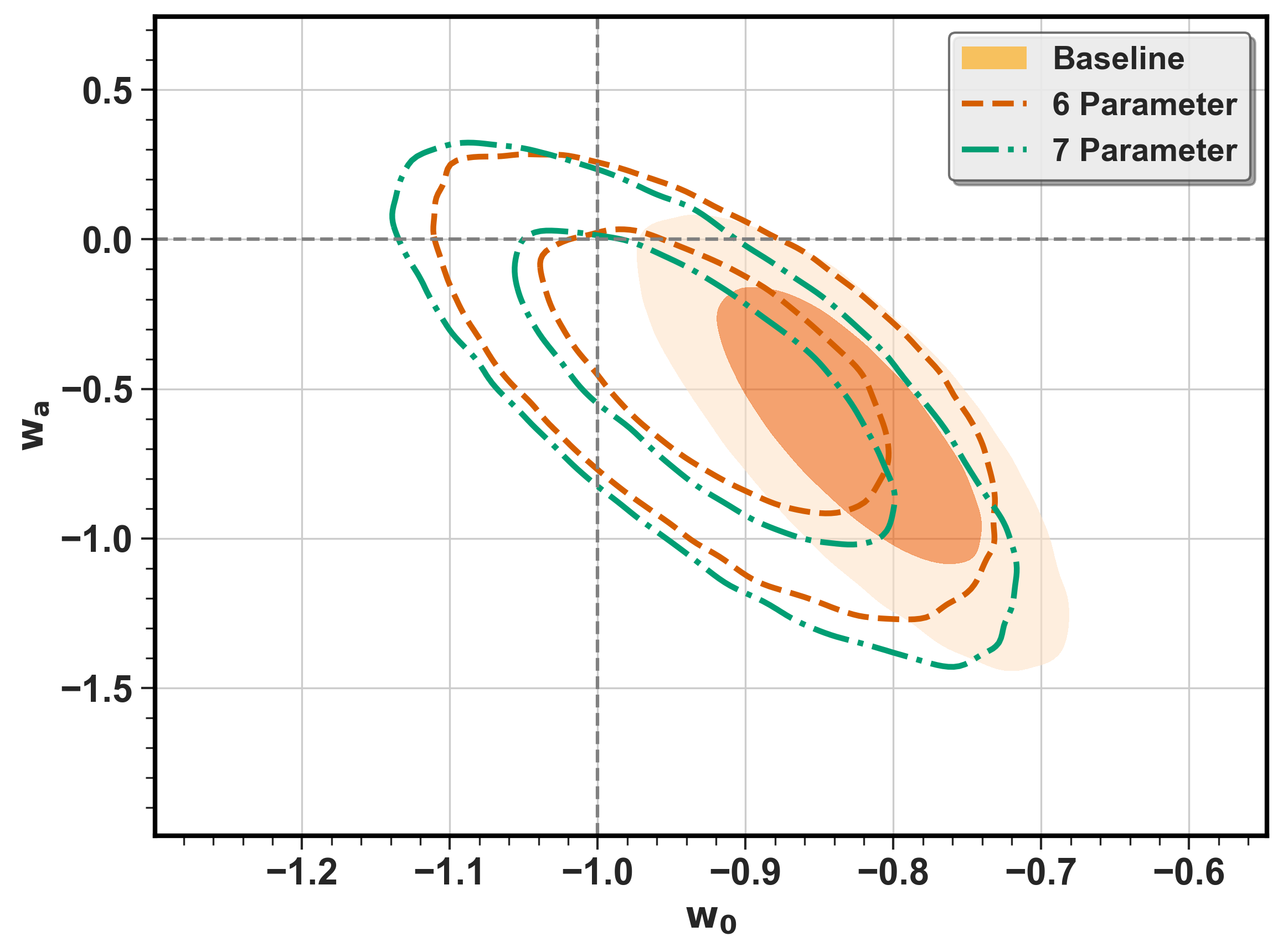}
    \caption{Contours in the dark energy EoS (EoS) parameters denoted by $(w_0,w_a)$ plane for three analyses: the Baseline fit (without correcting from mismatch in CDDR) (filled orange) constraining $\{H_0,\Omega_m,w_0,w_a\}$; the 6-parameter fit (dashed orange) adding $(d_0,d_1)$; and the 7-parameter fit (dash-dot green) further including $d_2$ (More details are given in Sec. \ref{sec:DataDrivenDDTest}). Allowing for distance duality deviations broadens the parameter space and shifts the best‐fit region toward the $\Lambda$CDM reference point $(w_0=-1,\;w_a=0)$ in agreement with previous cosmological results, highlighting the importance of consistency checks between distance indicators before combining datasets for cosmological inference.}
    \label{fig:ContourPlot}
\end{figure}

In Figure~\ref{fig:ContourPlot} we show the joint marginalized posteriors on \((w_0, w_a)\), adopting a Planck 2018 prior on \(\Omega_m\) \cite{Planck:2018vyg}.  The baseline four-parameter fit (filled orange) yields $(w_0, w_a)\approx(-0.83,\,-0.62)\,$. Allowing for CDDR violations by adding two parameters \((d_0,d_1)\) (six-parameter case; dashed orange) broadens the contours and shifts the peak to $(w_0, w_a)\approx(-0.92,\,-0.44)\,$. Introducing a third term \((d_2)\) (seven-parameter case; dash-dot green) moves the best-fit even closer to \(\Lambda\)CDM, $(w_0, w_a)\approx(-0.92,\,-0.49)\,$. The full numerical results are listed in Table~\ref{tab:Summary}. This progression toward \((-1,0)\) highlights how mild departures from the standard CDDR relation in the data can substantially alter dark energy constraints, and underscores the importance of consistency checks between Pantheon\(+\) and DESI BAO before combining them.  With these corrections in place, the DESI+Pantheon inference becomes fully consistent with a cosmological constant.

This paper is organized as follows. In Section \ref{sec:DDTest}, we describe the distance duality test used in our study to examine the consistency between various observational datasets. Section \ref{sec:DataDrivenDDTest} focuses on the Pantheon+ and DESI BAO data, assessing the mutual consistency of these datasets. In Section \ref{sec:method}, we detail the methodology employed to jointly estimate the cosmological parameters relevant to our analysis. Section \ref{sec:result} summarizes the results of this joint analysis, and Section \ref{sec:Discussion} explores their scientific implications. Finally, Section \ref{sec:Conclusion} offers a concise summary of our key findings and suggests avenues for future work.

\section{A Primer on the CDDR Test}
\label{sec:DDTest}

The CDDR is a fundamental prediction that emerges from photon number conservation in any metric theory of gravity, as articulated by Etherington’s reciprocity theorem. Independent of the details of the cosmological model, the CDDR establishes a connection between the luminosity distance, \(D_L(z)\), measured using standard candles (such as SNIa), and the angular diameter distance, \(D_A(z)\), determined using standard rulers (such as BAO) at a given redshift \(z\). This relation is expressed as
\begin{equation}
    D_L(z) = (1+z)^2\,D_A(z).
    \label{eq:DistDuality}
\end{equation}

When combining distance measurements from different observational probes, it is crucial to validate the CDDR. Each dataset carries its own systematic uncertainties, and if these are not properly accounted for, hidden biases may jeopardize the joint inference of cosmological parameters. For example, intergalactic dust attenuation, evolution in the properties of supernova progenitors, or even exotic physics could lead to deviations in the observed supernova luminosity distances, thus creating an apparent violation of the CDDR. Such discrepancies may either indicate unaccounted-for astrophysical systematic errors or point toward new physics. This CDDR test has also been done with gravitational waves sources with BAO for a model-independent propagation test of General Relativity as demonstrated in \citep{Afroz:2024joi,Afroz:2024oui,Afroz:2023ndy,Mukherjee:2020mha}.

To robustly test for these effects, we introduce a phenomenological distance duality coefficient, \(\mathcal{D}(z)\), defined by
\begin{equation}
    D_A^{\mathrm{obs}_1}(z) = \mathcal{D}(z)(1+z)^{-2}\,D_L^{\mathrm{obs}_2}(z),
\end{equation}
where \(D_A^{\mathrm{obs}_1}(z)\) is the angular diameter distance inferred from one type of observation and \(D_L^{\mathrm{obs}_2}(z)\) is the distance determined from another. Under ideal conditions, we expect \(\mathcal{D}(z) = 1\) at all redshifts. Any deviation from unity would signal inconsistencies between the distance measurements, discrepancies that need to be carefully addressed when combining datasets to ensure the integrity of joint cosmological parameter estimation. In principle, one could also perform this consistency test using a fiducial cosmological model by comparing one of the observed distances with the model-predicted distance.

\section{Data-Driven CDDR Test for supernovae samples and DESI DR2}
\label{sec:DataDrivenDDTest}

We combine the Pantheon$+$ SNIa data with DESI BAO measurements to evaluate whether the calibrated SNIa distances are consistent with the BAO-inferred distance scale. The distance duality coefficient is defined as
\begin{equation}
    D_{A}^{\rm DESI}(z)=\mathcal{D}(z)(1+z)^{-2}\,D_L^{\rm SNIa}(z),
    \label{eq:DDCoeff}
\end{equation}

where \(D_A^{\rm DESI}(z)\) is the distance derived from DESI BAO measurements using the CMB-calibrated sound horizon \(r_d = 147.09\,\mathrm{Mpc}\) \citep{Planck:2018vyg}, and \(D_L^{\rm SNIa}(z)\) is the distance directly inferred from the SNIa data. We can define a quantity $D_L^{\rm DESI}(z)\equiv  (1+z)^2D_A^{\rm DESI}$, where all the quantities on the right-hand side of the equation come from the data. In this analysis, we utilize the transverse BAO observable $D_M(z)/r_d$ from DESI DR2, which directly relates to the angular diameter distance via $D_M(z) = (1+z)D_A(z)$. The DESI DR2 data provide three types of compressed distance observables: the transverse measurement $D_M/r_d$ (which can be directly converted to $D_A$ after assuming an $r_d$ value), the radial measurement $D_H/r_d$ (which provides $H(z)$), and the spherically-averaged measurement $D_V/r_d$ (a combination of the angular and radial information).

\begin{figure}[ht]
    \centering
    \includegraphics[width= 8.0cm, height =6.0cm]{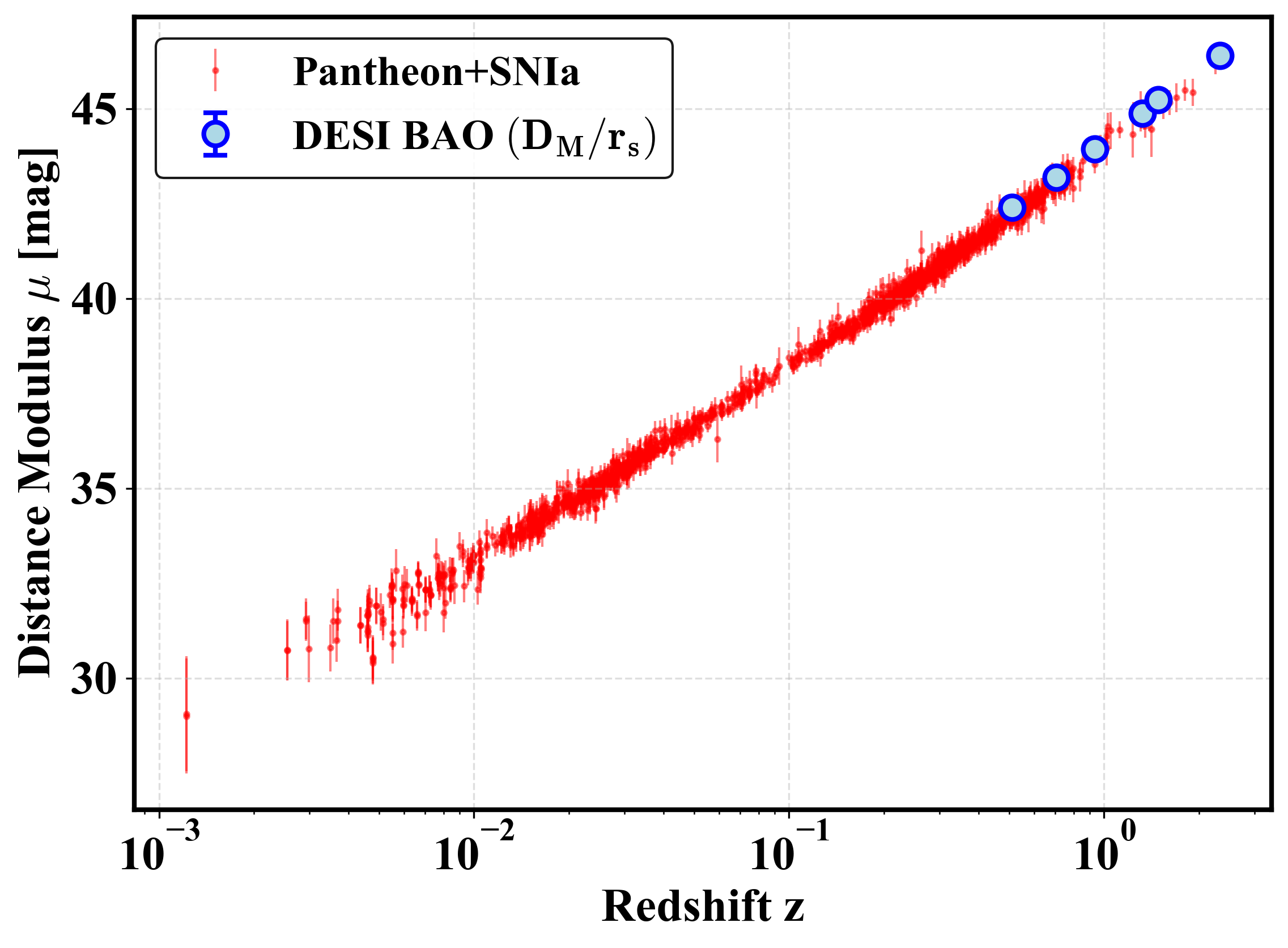}
    \caption{Comparison of cosmological distance measurements as a function of redshift, displayed as distance modulus $\mu$ [mag]. Red points show the Pantheon+ supernova sample (1701 SNIa), while blue points show DESI BAO measurements converted to distance modulus using the transverse observable $D_M/r_s$ with the fiducial sound horizon $r_d = 147.09$ Mpc from Planck 2018 \citep{Planck:2018vyg}. The conversion follows $\mu = 5\log_{10}[(1+z) (D_M / r_d) \times r_d] + 25$, where $D_L = (1+z)D_M = (1+z)^2 D_A$. Error bars represent 1$\sigma$ uncertainties; for DESI, these are derived from the diagonal elements of the covariance matrix. This visualization demonstrates both datasets used in our CDDR consistency analysis on a common distance scale, facilitating direct visual comparison between the two independent cosmological probes.}
    \label{fig:DataPlot}
\end{figure}

Figure~\ref{fig:DataPlot} displays both the DESI BAO and Pantheon+ SNIa measurements on a common distance modulus scale for direct visual comparison. For this visualization, we convert the DESI transverse BAO observable $D_M/r_s$ to distance modulus using $\mu = 5\log_{10}[(1+z)(D_M/r_d) \times r_d] + 25$\, where we adopt the fiducial sound horizon $r_d = 147.09$ Mpc from Planck 2018 \cite{Planck:2018vyg}. The Pantheon+ Type supernova sample provides distance moduli for 1701 SNIa over the range \(0.001 < z < 2.26\) along with the  DESI DRII BAO measurements.

We note that while alternative determinations of $r_d$ exist in the literature (e.g., $r_d = 147.5 \pm 2.0$ Mpc; $r_d = 136.4 \pm 3.5$ Mpc from Planck with SH0ES $H_0$ prior \citep{Planck:2018vyg}), our choice ensures direct comparability with the DESI DR2 published results. Although the cosmic distance duality relation $D_L = (1+z)^2 D_A$ is a geometric relation, its empirical test relies on angular diameter distances inferred from BAO measurements and therefore depends on the assumed value of $r_d$. As a result, variations in $r_d$ can affect the inferred distance-duality and cosmological parameters through their degeneracies with the overall distance scale. To explicitly assess this dependence, we repeat the full analysis using an alternative value of the sound horizon scale and present the results in Appendix~\ref{app:rd_dependence}.

\section{Methodology for Joint Analysis of SNIa and BAO Data}
\label{sec:method}

We base our analysis on datasets of SNIa, Pantheon with the measurements from BAO and CMB. In our analysis the comoving sound horizon is fixed to \(r_d = 147.09\) Mpc (from Planck \cite{Planck:2018vyg}), and we adopt the BAO likelihood as implemented in the \texttt{cobaya} \citep{2019asclsoft10019T} framework, where it is modeled by a multivariate Gaussian in the compressed parameters with their corresponding covariance matrices.

\begin{figure}[t]
    \centering
    \includegraphics[width=8.5cm, height=7.5cm]{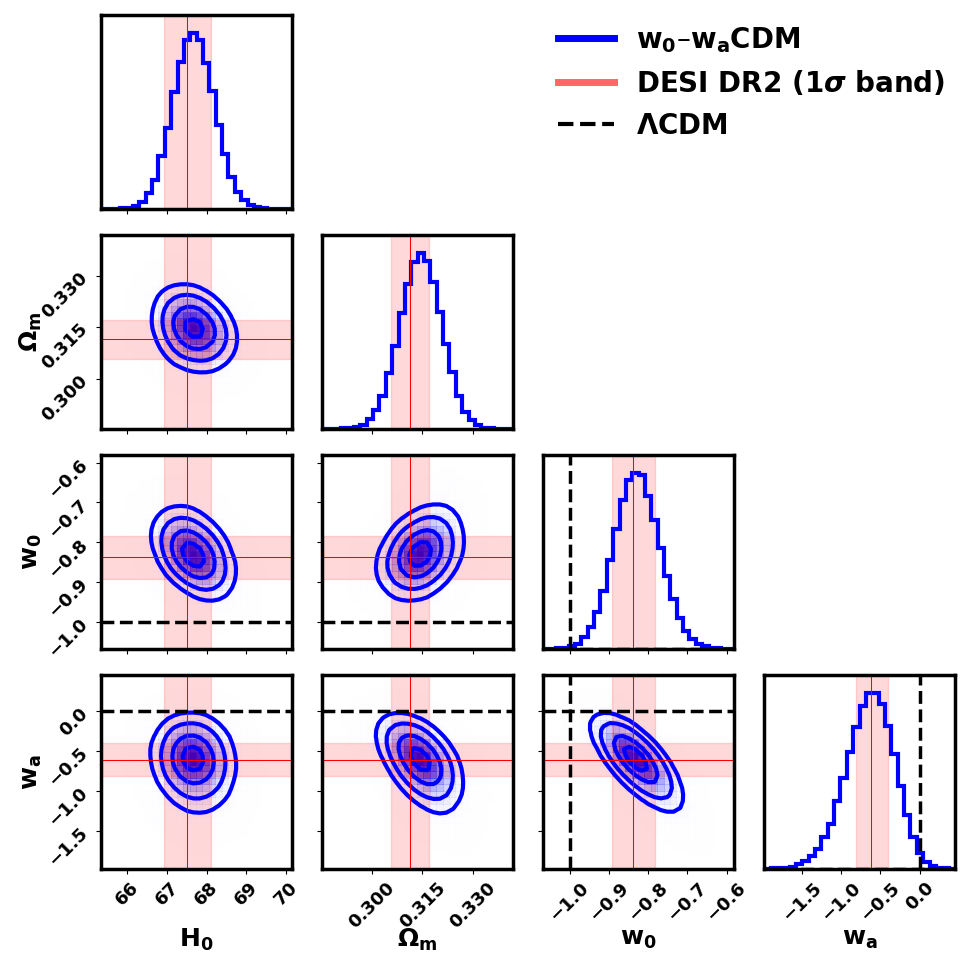}
    \caption{Baseline constraints on the parameter set \(\{H_0, \Omega_m, w_0, w_a\}\) in a flat \(w_0\text{-}w_a\)CDM model without including the distance duality parameters. The blue contours show the results of our analysis using DESI DR2 + Pantheon+ data, while the red shaded regions indicate the DESI DR2 (BAO + SNIa + Planck $\rm{\Omega_m}$ prior) \(1\sigma\) constraints. We impose a Gaussian Planck prior on \(\Omega_m\) (mean = 0.315, standard deviation = \(0.007\)), and adopt flat priors on \(H_0 \in [60, 80]\), \(w_0 \in [-2, 1]\), and \(w_a \in [-3, 3]\), together with the condition \(w_0 + w_a < 0\) to ensure viable past-light-cone histories. This plot shows that our results are in excellent agreement with the DESI DR2 constraints when no distance duality parameter is included.}
    \label{fig:Baseline}
\end{figure}

\subsection{Baseline Inference without the Distance Duality Coefficient in the likelihood}

In the \emph{baseline model}, we assume that the SNIa luminosity distances conform to a flat \(w_0w_a\)CDM cosmology without any redshift-dependent modification. Concretely, the theoretical SNIa distance modulus is computed as
\begin{equation}
\mu_{\rm th}(z) \;=\; 5\log_{10}\!\left[\frac{D_L^{\rm fid}(z)}{10\,\mathrm{pc}}\right],
\end{equation}
where the fiducial luminosity distance \(D_L^{\rm fid}(z)\) is evaluated using a \(w_0\text{-}w_a\)CDM cosmology with the following four parameters \(\{H_0,\,\Omega_m,\,w_0,\,w_a\}\). The corresponding SNIa Pantheon+  likelihood takes the form
\begin{align}
-2\ln\mathcal{L}_{\rm SNIa} \; & =\; \sum_{i,j}\Bigl[\mu_{\rm obs}(z_i) - \mu_{\rm fid}(z_i)\Bigr]\, \nonumber \\ & \times 
C_{ij}^{-1}\,\Bigl[\mu_{\rm obs}(z_j) - \mu_{\rm fid}(z_j)\Bigr],
\end{align}
where \(\mu_{\rm obs}(z_i)\) denotes the observed distance modulus and \(C_{ij}\) is the full covariance matrix (including both statistical and systematic uncertainties).

In practice, because the absolute magnitude \(M\) (which sets the overall normalization of \(\mu\)) is unknown {\em apriori}, we promote it to a free nuisance parameter and marginalize over it analytically.  Concretely, this involves subtracting a single, unknown constant offset-corresponding to \(M\)-from each residual, assigning that offset a flat prior, and carrying out the Gaussian integral in closed form.  The resulting likelihood is algebraically equivalent to using residuals re-centered by the best‐fit offset, ensuring that the supernovae constrain only relative distances and do not independently determine the absolute distance scale.  This approach follows the Pantheon\,+ methodology exactly, and guarantees that our SNIa likelihood is fully marginalized over \(M\).

The DESI BAO likelihood is implemented within the \texttt{cobaya} framework using CAMB \citep{Lewis:1999bs} to compute the theoretical BAO observables. computes the theoretical values for each BAO observable (i.e., \(D_M/r_s\), \(D_H/r_s\), and \(D_V/r_s\)) and then compares them to the observed BAO data using the inverse covariance matrix. The corresponding log-likelihood, \(-2\ln\mathcal{L}_{\rm BAO}\), is therefore given by
\begin{align}
-2\ln\mathcal{L}_{\rm BAO} \;& =\; \sum_{i,j}\Bigl[\mathcal{D}_{\rm obs}(z_i) - \mathcal{D}_{\rm th}(z_i)\Bigr]\, \nonumber \\& \times
(C^{-1}_{\rm BAO})_{ij}\, \Bigl[\mathcal{D}_{\rm obs}(z_j) - \mathcal{D}_{\rm th}(z_j)\Bigr],
\end{align}
where the theoretical predictions \(\mathcal{D}_{\rm th}(z)\) are a function of \(\{H_0,\,\Omega_m,\,w_0,\,w_a\}\) through the background evolution computed by CAMB. The joint likelihood is then constructed by combining the SNIa and BAO likelihoods,
\begin{equation}
\ln\mathcal{L}_{\rm joint} \;=\; \ln\mathcal{L}_{\rm SNIa} + \ln\mathcal{L}_{\rm BAO},
\end{equation}
which is used to constrain the parameter set \(\{H_0,\,\Omega_m,\,w_0,\,w_a\}\).

\begin{figure*}[t]
    \centering
    \includegraphics[width=16cm, height=15cm]{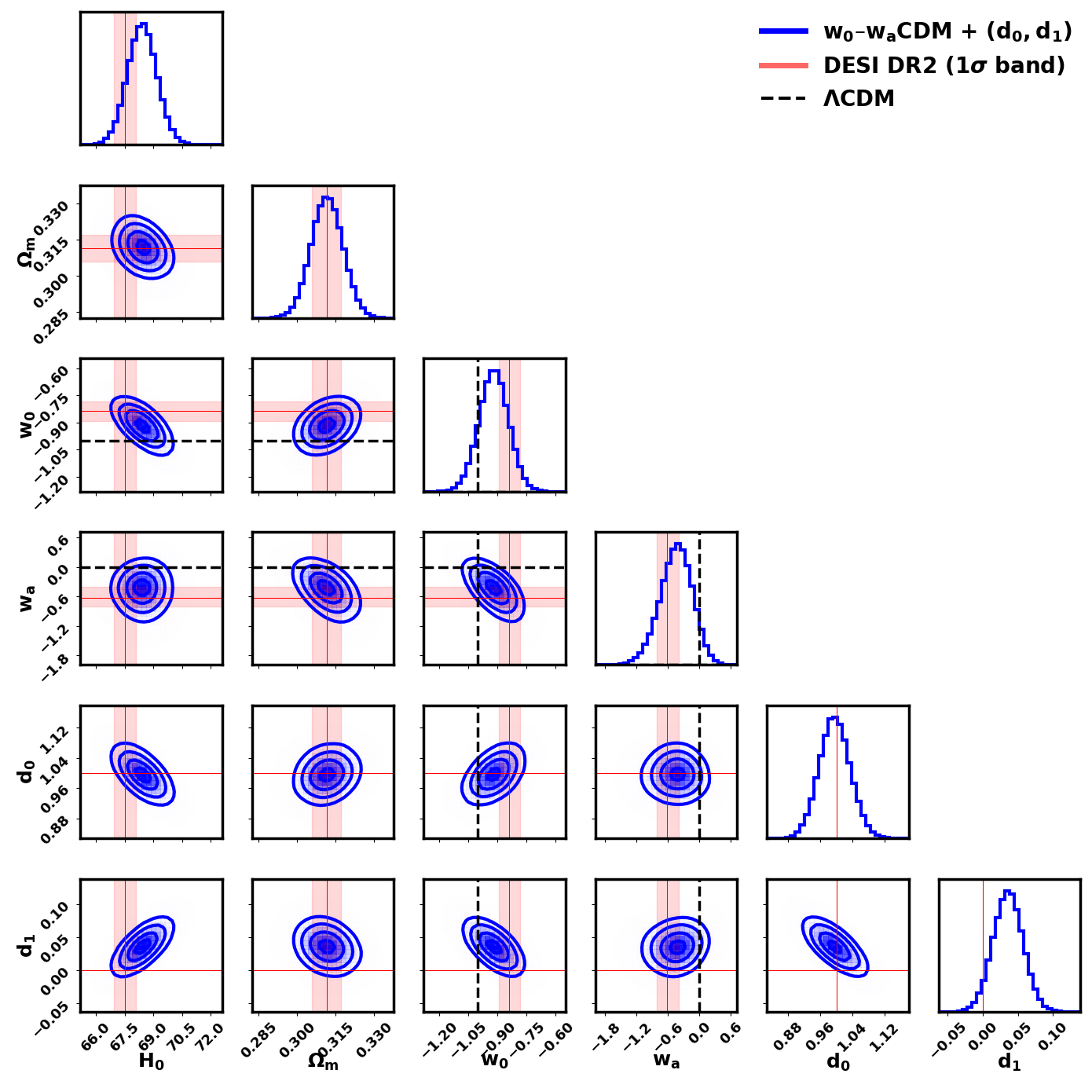} 
    \caption{Joint six‐parameter posterior constraints on the parameter set \(\{H_0,\Omega_m,w_0,w_a,d_0,d_1\}\) in a flat \(w_0\text{-}w_a\)CDM model including distance-duality coefficients. The blue contours represent the joint constraints from DESI DR2 + Pantheon+ for our extended framework, while the red shaded regions correspond to the DESI DR2 (BAO + SNIa + Planck $\rm{\Omega_m}$ prior) results. We impose a Gaussian Planck prior on \(\Omega_m\) (mean = 0.315, $\sigma$ = 0.007), flat priors on \(H_0\in[60,80]\), \(w_0\in[-2,1]\) and \(w_a\in[-3,3]\) with the requirement \(w_0 + w_a < 0\), and flat priors on the distance-duality coefficients \(d_0\in[-1,2]\) and \(d_1\in[-1,1]\).}
    \label{fig:6param}
\end{figure*}

\subsection{Extended Inference with a Distance Duality Coefficient}

To probe the effect of variation of duality coefficient on the cosmological inference, we incorporate the distance duality coefficient \(\mathcal{D}(z)\) into the SNIa likelihood. The inclusion of \(\mathcal{D}(z)\) in the likelihood provides a flexible framework for investigating the degeneracy between the distance calibration and the underlying cosmological parameters. In particular any systematic variation in \(\mathcal{D}(z)\) can be partially degenerate with shifts in parameters such as \(H_0\) or the dark energy EoS parameters \((w_0,\,w_a)\). This structure allows us to diagnose how much of the apparent tension between the SNIa and BAO measurements could be attributed to a breakdown in the standard distance duality relation. Specifically, we define the modified theoretical luminosity distance as
\begin{equation}
D_L^{\rm mod}(z) \;\equiv\; \mathcal{D}(z)\,D_L^{\rm fid}(z).
\end{equation}
Accordingly, the modified thoeretical SNIa distance modulus becomes
\begin{equation}
\mu_{\rm mod}(z) \;=\; \mu_{\rm fid}(z) + 5\,\log_{10}\bigl[\mathcal{D}(z)\bigr].
\end{equation}
This modification leads to the following form for the SNIa likelihood: 

\begin{align}
-2\ln\mathcal{L}_{\rm SNIa}^{\rm Modified} \; & =\; \sum_{i,j}\Bigl[\mu_{\rm obs}(z_i) - \mu_{\rm fid}(z_i) - 5\,\log_{10}\bigl(\mathcal{D}(z_i)\bigr)\Bigr]\, \nonumber \\ & \times 
C_{ij}^{-1}\,\Bigl[\mu_{\rm obs}(z_j) - \mu_{\rm fid}(z_j) - 5\,\log_{10}\bigl(\mathcal{D}(z_j)\bigr)\Bigr].
\end{align}

Since \(\mathcal{D}(z)\) modifies only the SNIa distances, the BAO likelihood remains unchanged. We thus construct the modified joint likelihood as the product of the SNIa modified and BAO likelihoods,
\begin{equation}
\mathcal{L}_{\rm joint}^{\rm Modified} \;=\; \mathcal{L}_{\rm SNIa}^{\rm Modified}\times\mathcal{L}_{\rm BAO}\,.
\end{equation}
The introduction of $\mathcal{D}(z)$ enables the model to detect any redshift-dependent tension or relative miscalibration between the SNIa and BAO datasets. Specifically, in the joint inference, $\mathcal{D}(z)$ is simultaneously constrained along with the cosmological parameters that govern the expansion history and distance--redshift relation. Any inconsistency in the absolute calibration or redshift evolution of SNIa distances relative to the BAO scale would lead to a compensating deviation in $\mathcal{D}(z)$, correlated with shifts in the inferred cosmological parameters. As a result, this formalism naturally takes care of redshift dependent calibration mismatch between two different distance probes BAO and SNe.

Using this joint modified likelihood, we simultaneously fit the cosmological parameters 
\(\{H_0,\;\Omega_m,\;w_0,\;w_a\}\) together with the distance‐duality scaling \(\mathcal{D}(z)\).  
We consider two parameterizations of \(\mathcal{D}(z)\):
\begin{align}
  \mathcal{D}(z) &= d_0 + d_1\,(1+z)\,, \label{eq:linearDD}\\
  \mathcal{D}(z) &= d_0 + d_1\,(1+z) + d_2\,(1+z)^2\,. \label{eq:quadraticDD}
\end{align}
In the standard distance‐duality scenario, one has \(d_0=1\) and \(d_1=d_2=0\), so that \(\mathcal{D}(z)=1\) at all redshifts. We choose polynomial parameterizations in \((1+z)\) for several reasons. First, they provide a natural Taylor expansion around the present epoch (\(z=0\)), making them well‐suited for capturing low‐to‐intermediate redshift systematics that may affect either the supernova or BAO datasets. Second, the coefficients have clear physical interpretation: \(d_0\) represents an overall normalization offset or calibration difference between the two distance measures, \(d_1\) captures a linear redshift evolution in the systematic effect (such as evolving supernova properties or calibration drift), and \(d_2\) describes higher‐order curvature or acceleration in the deviation.

We therefore perform:
\begin{itemize}
  \item a six‐parameter analysis \(\{H_0,\Omega_m,w_0,w_a,d_0,d_1\}\) using Eq.~\eqref{eq:linearDD}, and  
  \item a seven‐parameter analysis \(\{H_0,\Omega_m,w_0,w_a,d_0,d_1,d_2\}\) using Eq.~\eqref{eq:quadraticDD},
\end{itemize}
employing the combined DESI DR2 and Pantheon\(+\) dataset.

To verify that our conclusions are not sensitive to this specific choice of functional form, we also test alternative parameterizations (exponential and logarithmic forms) in Appendix~\ref{app:Alternative_Dz}. In all cases, we find consistent qualitative behavior: the inclusion of $\mathcal{D}(z)$ systematically shifts the inferred dark energy EoS parameters $(w_0, w_a)$ toward the $\Lambda$CDM values $(-1, 0)$ relative to the baseline analysis without distance-duality corrections. Importantly, the inferred constraints on $H_0$ and $\Omega_m$, as well as the direction and magnitude of the shifts in $(w_0, w_a)$, remain stable across all parameterizations, demonstrating that our results are robust against the functional form adopted for $\mathcal{D}(z)$

\begin{figure*}[t]
    \centering
    \includegraphics[width=16cm, height=16cm]{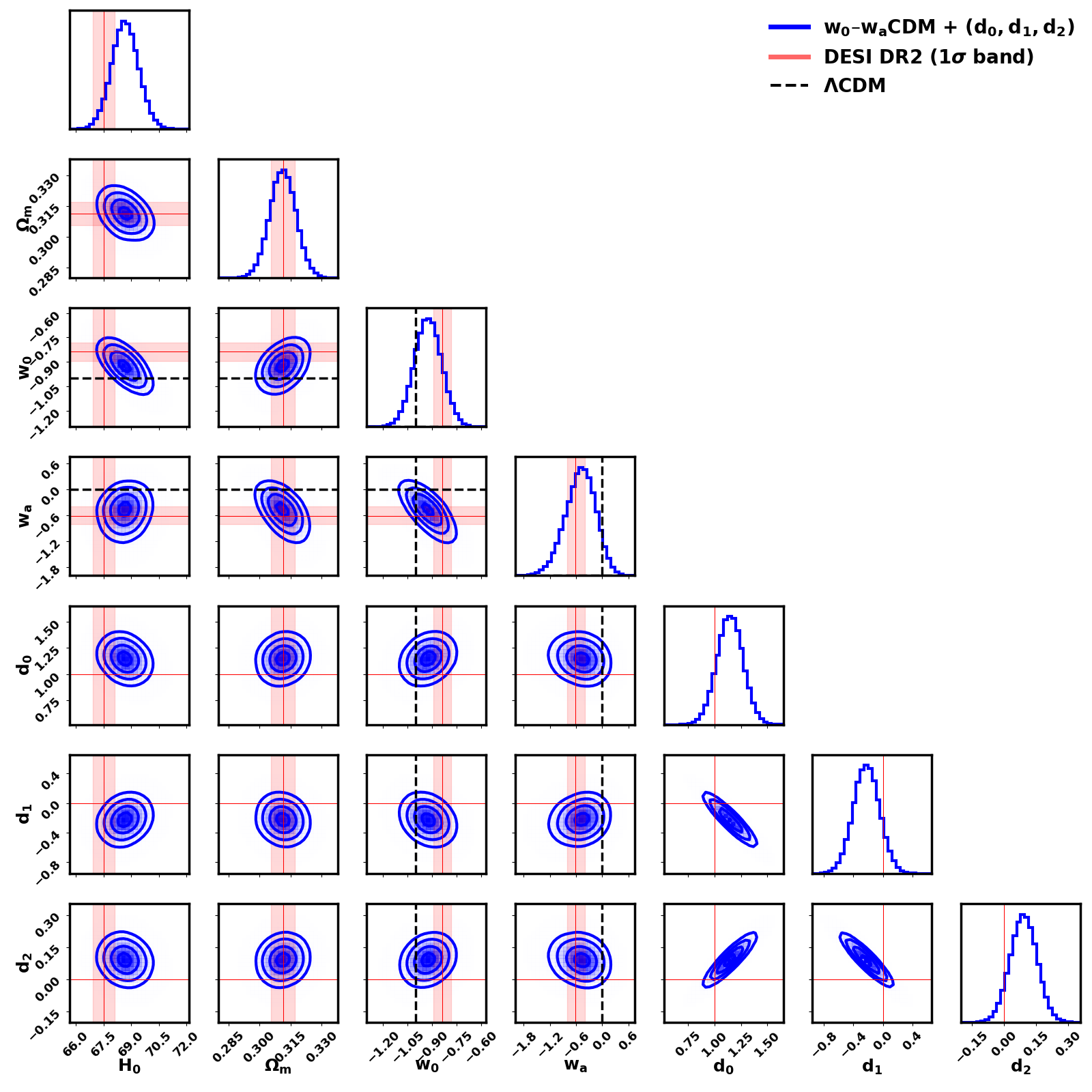} 
    \caption{Joint seven‐parameter posterior constraints on the parameter set \(\{H_0,\Omega_m,w_0,w_a,d_0,d_1,d_2\}\) in a flat \(w_0\text{-}w_a\)CDM model including distance duality coefficients. The blue contours represent the joint constraints from DESI DR2 + Pantheon+ for our extended framework, while the red shaded regions correspond to the DESI DR2 (BAO + SNIa + Planck $\rm{\Omega_m}$ prior) results. We impose a Gaussian Planck prior on \(\Omega_m\) (mean = 0.315, $\sigma$ = 0.007), flat priors on \(H_0\in[60,80]\), \(w_0\in[-2,1]\) and \(w_a\in[-3,3]\) with the requirement \(w_0 + w_a < 0\), and flat priors on the distance duality coefficients \(d_0\in[-1,2]\), \(d_1\in[-1,1]\) and \(d_2\in[-1,1]\).}
    \label{fig:7param}
\end{figure*}

\section{Results}
\label{sec:result}
\subsection{Baseline case: No inclusion of distance-duality coefficient in the likelihood}

In Figure~\ref{fig:Baseline}, we present the posterior distributions for the baseline model obtained by combining the standard Pantheon\(+\) SNIa and DESI BAO likelihoods within a fully Bayesian framework, adopting a Gaussian Planck prior on \(\Omega_m\) (mean 0.315, \(\sigma=0.007\)) and flat priors \(H_0\in[60,80]\), \(w_0\in[-2,1]\), and \(w_a\in[-3,3]\) subject to \(w_0+w_a<0\). Notably, whereas the SNIa data tend to pull the Hubble constant \(H_0\) toward higher values, the BAO measurements favor slightly lower values, resulting in dark energy parameters that depart from the canonical \((-1,0)\). In particular, \(w_0\) shifts mildly away from \(-1\) and \(w_a\) indicates a slight evolution in the dark energy EoS.

\subsection{Extended case: Inclusion of distance-duality coefficient in the likelihood}

Figures~\ref{fig:6param} and \ref{fig:7param} display the joint six‐ and seven‐parameter posterior distributions, respectively, obtained from a fully Bayesian analysis combining DESI DR2 and Pantheon\(+\) likelihoods with the modified SNIa model.  We impose:
\begin{itemize}
  \item A Gaussian Planck prior on \(\Omega_m\) (mean = \(0.315\), standard deviation \(=0.007\)),
  \item Flat priors \(H_0\in[60,80]\), \(w_0\in[-2,1]\), \(w_a\in[-3,3]\) with \(w_0+w_a<0\)\footnote{This is imposed to obtain the results to be consistent with DESI DR2 \cite{DESI:2025zgx}},
  \item Flat priors on the distance-duality coefficients \(d_0\in[-1,2]\), \(d_1\in[-1,1]\), and (for the seven‐parameter case) \(d_2\in[-1,1]\).
\end{itemize}

In the six‐parameter analysis (Figure~\ref{fig:6param}), introducing \(d_0\) and \(d_1\) broadens the \(w_0\text{-}w_a\) contours relative to the baseline and shifts their peak closer to the \(\Lambda\)CDM point \((w_0=-1,\;w_a=0)\).  The overall normalization \(d_0\) remains anchored near unity, while the linear evolution term \(d_1\) shows only a mild redshift dependence. 
In the seven‐parameter analysis (Figure~\ref{fig:7param}), adding the quadratic coefficient \(d_2\) further relaxes the dark‐energy constraints, introducing  degeneracies among \(\{d_0,d_1,d_2\}\).  The quadratic term \(d_2\) itself mildly deviates from zero, and \(d_0\) and \(d_1\) exhibit trends similar to the six‐parameter case.

An important point to notice from Figures ~\ref{fig:6param} and ~\ref{fig:7param} that there exists strong correlations between the cosmological parameters and the distance-duality parameters ($d_0,\, d_1,\, d_2$). Due to the existence of these correlation, the cosmological parameters can be driven towards an incorrect value, if there exists non-zero contribution from $d_0,\, d_1,\, d_2$. As a result joint-estimation of the cosmological parameters along with the distance-duality coefficients can reveal possible systematic and can mitigate its influence on the cosmological parameters. It is important to note here that the correlation of the parameter $d_2$ with other parameters, drives the value of $w_0$ and $w_a$ towards cosmological constant, and peaks a little away from the $d_2=0$ indicating hint towards a redshift dependent mismatch between the two datasets, which can mimic redshift evolving dark energy. Though the posterior distribution of none of the parameters ($d_0$, $d_1$, and $d_2$) are not statistically significantly away from the expected values (shown by the vertical line), their about 1-$\sigma$ shifts points towards some possible redshift dependent contamination.

In addition, we assess the robustness of these results against both the assumed sound horizon scale and the functional form of the distance-duality coefficient. Repeating the six- and seven-parameter analyses using alternative values of the sound horizon $r_d$ as well as logarithmic and exponential parameterizations of $\mathcal{D}(z)$, we find consistent qualitative behavior in all cases. Variations in $r_d$ primarily rescale the inferred Hubble constant, in accordance with the inverse relation $H_0 \propto r_d^{-1}$, while leaving the matter density largely unaffected. Owing to the partial degeneracy between $r_d$ and the distance-duality parameters, changes in the sound horizon can be partially absorbed by shifts in $\mathcal{D}(z)$, but the correlations between the cosmological and distance-duality parameters persist across all parameterizations. Importantly, irrespective of the choice of $\mathcal{D}(z)$ or $r_d$, the inclusion of distance-duality corrections systematically shifts the dark energy EoS parameters $(w_0,w_a)$ toward the $\Lambda$CDM values.

\subsection{Summary of all the cases}

In this section, we present quantitative results obtained from the combined DESI DR2 and SNIa Pantheon\(+\) datasets under three different modeling assumptions, each progressively introducing additional freedom related to the distance-duality relation. Specifically, we first analyze the baseline \(w_0\text{-}w_a\text{CDM}\) model (varying only the four cosmological parameters \(\{H_0,\Omega_m,w_0,w_a\}\)), then extend the analysis to include two distance-duality coefficients \((d_0,d_1)\), and finally further extend to a seven-parameter scenario that incorporates an additional quadratic term \(d_2\). These incremental extensions enable us to systematically assess how relaxing the standard assumption of distance duality impacts cosmological parameter inference. The full results for all parameters in each model variant are summarized in Table~\ref{tab:Summary}.

\begin{table*}[ht]
\centering
\footnotesize
\setlength{\tabcolsep}{3pt}
\small 
\renewcommand{\arraystretch}{1.5} 
\resizebox{\textwidth}{!}{%
  \begin{tabular}{|l|c|c|c|c|c|c|c|c|c|}
    \hline
    \multicolumn{10}{|c|}{\textbf{Summary of the Constraints on the cosmological parameters and $\mathcal{D}(z)$ parameters.}} \\
    \hline
    Model & $H_0$ & $\Omega_m$ & $w_0$ & $w_a$ & $d_0$ & $d_1$ & $d_2$ & $\chi_\nu^2$ & $\text{BF}$ \\
    \hline
    $w_0\text{-}w_a \text{CDM}$ & $67.67^{+0.51}_{-0.51}$ & $0.31^{+0.01}_{-0.01}$ & $-0.83^{+0.06}_{-0.06}$ & $-0.62^{+0.29}_{-0.32}$ & \text{-} & \text{-} & \text{-} & $1.04$ & $2.47$ \\
    \hline
    \shortstack{$w_0\text{-}w_a \text{CDM}$\\$+(d_0,\,d_1)$} & $68.43^{+0.76}_{-0.76}$ & $0.31^{+0.01}_{-0.01}$ & $-0.92^{+0.08}_{-0.08}$ & $-0.44^{+0.29}_{-0.32}$ & $1.00^{+0.04}_{-0.04}$ & $0.04^{+0.02}_{-0.02}$ & \text{-} & $0.68$ & $54.14$ \\
    \hline
    \shortstack{$w_0\text{-}w_a \text{CDM}$\\$+(d_0,\,d_1,\,d_2)$} & $68.66^{+0.73}_{-0.72}$ & $0.31^{+0.01}_{-0.01}$ & $-0.92^{+0.08}_{-0.08}$ & $-0.49^{+0.33}_{-0.36}$ & $1.15^{+0.12}_{-0.12}$ & $-0.22^{+0.18}_{-0.17}$ & $0.09^{+0.06}_{-0.06}$ & $1.06$ & $49.47$ \\
    \hline
  \end{tabular}
}
\caption{Summary of the measured values for the Hubble constant $H_0$, matter density $\Omega_m$, the dark energy EoS parameters $(w_0, w_a)$, and the distance duality coefficients $(d_0, d_1, d_2)$. Results are shown for three successive model variants: the baseline $w_0\text{-}w_a$CDM model without any distance duality parameters, the extension including two coefficients $(d_0, d_1)$, and the full extension with all three coefficients $(d_0, d_1, d_2)$. The reduced chi-square values ($\chi_\nu^2$) are computed at the best-fit parameters for each model. The Bayes factors ($\text{BF}$) are computed using the Savage-Dickey density ratio at the fiducial $\Lambda$CDM point $(w_0 = -1, w_a = 0)$ with uniform priors over $w_0 \in [-2,1]$ and $w_a \in [-3,3]$.}
\label{tab:Summary}
\end{table*}

In addition to constraining the model parameters, we evaluate the fit quality using the reduced chi-square values ($\boldsymbol{\chi^2_\nu}$) for each model variant at their respective best-fit parameters. The \textbf{baseline} $\mathbf{w_0\text{-}w_a}$\textbf{CDM} model yields $\mathbf{\boldsymbol{\chi^2_\nu} = 1.04}$. Introducing a redshift-dependent linear correction, the $\mathbf{w_0\text{-}w_a}$\textbf{CDM}$\mathbf{+(d_0, d_1)}$ model reduces the chi-square to $\mathbf{\boldsymbol{\chi^2_\nu} = 0.68}$. Extending the correction to a quadratic form, the $\mathbf{w_0\text{-}w_a}$\textbf{CDM}$\mathbf{+(d_0, d_1, d_2)}$ model yields $\mathbf{\boldsymbol{\chi^2_\nu} = 1.06}$. 

To further assess model preference, we compute the Bayes factors using the Savage\text{-}Dickey density ratio \citep{dickey1971weighted, Trotta:2008qt}, which provides an exact and computationally efficient method for evaluating Bayes factors when comparing nested models. In this case, the $\Lambda$CDM model corresponds to a special point in the extended $w_0$--$w_a$ parameter space, defined by $w_0=-1$ and $w_a=0$, and the Savage--Dickey ratio allows the Bayes factor to be obtained directly from the ratio of the posterior to prior densities evaluated at these fiducial values. These Bayes factors are evaluated at the fiducial values $w_0 = -1$ and $w_a = 0$, assuming uniform priors over the ranges $w_0 \in [-2, 1]$ and $w_a \in [-3, 3]$. For the baseline $\mathbf{w_0\text{-}w_a}$\textbf{CDM} model, the Bayes factor at this fiducial point is $\mathbf{2.47}$. For the extended $\mathbf{w_0\text{-}w_a}$\textbf{CDM}$+\mathbf{(d_0, d_1)}$ model, the Bayes factor increases to $\mathbf{54.14}$, and for the full $\mathbf{w_0\text{-}w_a}$CDM$+\mathbf{(d_0, d_1, d_2)}$ model, it is $\mathbf{49.47}$. This trend indicates that models incorporating distance duality corrections tend to produce posterior distributions more tightly clustered around the $\Lambda$\textbf{CDM} values $(w_0 = -1, w_a = 0)$, resulting in a higher posterior density at the fiducial point and consequently larger Bayes factors via the Savage\text{-}Dickey ratio. In contrast, the baseline DESI-only constraints favor values of $w_0$ and $w_a$ that deviate from $\Lambda$CDM. It is important to emphasize that Bayes factors are sensitive to both the prior choice and the shape of the posterior distribution. Since all models are evaluated using the same uniform priors, the higher Bayes factors in the extended models arise from their posteriors being more sharply peaked at the fiducial $\Lambda$CDM point, rather than from differences in prior volume.

In Figure~\ref{fig:ContourPlot}, we present the joint marginalized posterior distributions in the \((w_0, w_a)\) plane, derived using the combined DESI DR2 and Pantheon+ datasets under different modeling assumptions about the distance-duality relation. The baseline analysis (filled orange contours) constrains the parameters \(\{H_0, \Omega_m, w_0, w_a\}\) without any corrections to distance duality. Here, the posterior is relatively tight and centered near \((w_0, w_a)=(-0.83^{+0.06}_{-0.06},-0.62^{+0.29}_{-0.32})\), slightly offset from the canonical \(\Lambda\)CDM point \((-1,0)\), agreeing with the Pantheon+ SNIa and DESI BAO results \cite{DESI:2025zgx}.

However, when we extend our analysis to allow for deviations in the CDDR by introducing additional parameters \((d_0, d_1)\) (the six-parameter scenario, dashed orange contours), we observe a clear shift in the central value of the $w_0$ and $w_a$ parameters toward \((w_0, w_a)=(-0.92^{+0.08}_{-0.08},-0.44^{+0.29}_{-0.32})\), closer to the \(\Lambda\)CDM prediction and also broadening of the error-bar. This indicates that the observed tension between datasets may partially (or completely) driven the signature of evolution of the dark energy EoS. Furthermore, the introduction of an additional quadratic parameter \(d_2\) (the seven-parameter scenario, green dash-dot contours) broadens the constraints even further, slightly altering the posterior peak to approximately \((w_0, w_a)=(-0.92^{+0.08}_{-0.08},-0.49^{+0.33}_{-0.36})\). Although the quadratic coefficient \(d_2\) itself remains consistent with zero, its inclusion significantly relaxes the dark energy constraints and highlights additional degeneracies among the parameters, as shown by the increasingly elongated shape of the contours.

To test the robustness of these conclusions, we further explored two potential sources of modeling uncertainty in Appendices~\ref{app:Alternative_Dz} and~\ref{app:rd_dependence}. In Appendix~\ref{app:Alternative_Dz}, we repeated the six-parameter analyses using alternative functional forms of the distance-duality coefficient, namely logarithmic and exponential parameterizations. In all cases, we find qualitatively consistent behavior: allowing deviations from the standard distance-duality relation systematically shifts the inferred dark energy EoS parameters $(w_0,w_a)$ toward their $\Lambda$CDM values, although the precise width and orientation of the contours vary due to parameter degeneracies. This demonstrates that the observed trend is not an artifact of a specific choice of $\mathcal{D}(z)$ parameterization, but a conclusion which is based on existing datasets.

Furthermore in Appendix~\ref{app:rd_dependence}, we assess the impact of the assumed sound horizon scale by repeating the full analysis with an alternative value of $r_d$. As expected, variations in $r_d$ primarily rescale the absolute distance scale and therefore lead to a systematic shift in the inferred Hubble constant through the relation $H_0 \propto r_d^{-1}$. Owing to degeneracies between $r_d$ and the distance-duality parameters, part of this rescaling can be absorbed by $\mathcal{D}(z)$, while the matter density remains largely unaffected. Importantly, despite these shifts, the qualitative impact of distance-duality corrections on the dark energy sector remains unchanged: the inclusion of $\mathcal{D}(z)$ consistently drives $(w_0,w_a)$ closer to the $\Lambda$CDM values. However, this analysis indicates an important effect of possible systematics due to the assumption on the value of sound horizon and tests the robustness of dark energy EoS inference. We have discussed these points of possible systematics in the following section.

These results clearly demonstrate that inconsistent datasets can substantially affect the inferred dark energy EoS parameters. In particular, the additional freedom allowed by the CDDR consistency reduces the apparent mild tension between Pantheon+ and DESI BAO data and significantly weakens the claim of evolving dark energy EoS. This work demonstrates the signatures of possible inconsistencies in the datasets and indicate that the current DESI DR2 with Pantheon+ results are likely to be impacted by this. This makes their analyses precise (due to combining large volume of data), but inaccurate in the inference of cosmological parameters. In future, CDDR tests proposed here, will be important to carry out on different datasets to test for consistency between different datasets.

\section{Discussion on the scientific implication of our findings}
\label{sec:Discussion}

A rigorous consistency test between independent cosmological probes is essential for precise and reliable parameter estimation. When combining datasets each affected by its own systematic uncertainties hidden biases can lead to erroneous joint inferences. For example, unrecognized redshift-dependent systematics in SNIa may distort the inferred luminosity distances, potentially yielding a spurious evolution in the dark energy EoS. It is equally important to consider that BAO measurements are not immune to systematic errors. DESI BAO data, while robust and internally consistent, can be subject to uncertainties in the determination of the comoving sound horizon \(r_d\) and residual systematics in the large-scale clustering analysis. These BAO specific uncertainties can further complicate the comparison of distance scales derived from different probes. Therefore, verifying the consistency of distance measures from SNIa and BAO is crucial to avoid biases in the joint cosmological inference. Such consistency tests ensure that the combined constraints reliably reflect the true underlying cosmology, rather than being driven by mismatched systematics inherent in any single dataset. This approach is especially important for future analyses that will merge supernova data with large-scale structure measurements, where the sensitivity to even subtle systematic effects will be significantly enhanced. The introduction of a redshift-dependent distance-duality coefficient $\mathcal{D}(z)$ has proven effective in capturing the possible systematics in the SNIa Pantheon+ dataset and the DESI BAO dataset. We interpret this result in two possible ways, each with important implications:

\textbf{Astrophysical Systematics in SNIa:} The need for $\mathcal{D}(z)\neq1$ may indicate residual systematics in SNIa standardization or calibration. Pantheon+ is a meticulously standardized sample, yet combining dozens of surveys and applying empirical bias corrections could leave small redshift-dependent offsets. In our analysis, $\mathcal{D}(z)$ effectively models a smooth version of such an offset.

\emph{Evolution of SNIa properties:} If the average properties of SNIa progenitors (e.g. metallicity, delay time) shift with redshift, the standardized luminosity might slowly drift. For instance, higher-$z$ SNIa occur in lower-metallicity environments on average, which could make them slightly fainter. Such effects are expected to be small, but a few hundredths of a magnitude over $\Delta z \sim 1$ is not implausible \citep{2010ApJ...711L..66B,2011MNRAS.414.1592B,Riess:2006yk,Panagia:2007eg,Jones:2013dta}.

\emph{Unaccounted dust contamination:} While the SNIa light-curve fits correct for color, a form of extinction that is wavelength-gray (or a circumgalactic/intergalactic component not fully captured by color corrections) could dim distant SNIa more than nearby ones. This could manifest as an $\mathcal{D}(z)>$1 \citep{2007A&A...464..465R,Goobar:2018smm,Vavrycuk:2019czf,Ostman:2004eh,Croft:2000ns}.

\emph{Calibration drift:} Combining many SNIa surveys (some at low $z$, some at high $z$) relies on calibrations which are associated with suystematic uncertainties \citep{Brout:2021mpj,Brownsberger:2021uue}. SNIa dataset has gone through several rigorous valaidation, more recently including JWST observations \cite{Riess:2024vfa}. However, presence of any non-zero systematic calibration error between low-$z$ and high-$z$ samples could appear as a distance modulus shift. 

\emph{Selection biases:} Malmquist bias (brightness selection effects) and survey strategy differences can cause subtle redshift-dependent biases in the observed SNIa population. The Pantheon+ analysis includes corrections for selection bias as a function of redshift; however, if these corrections are slightly mis-estimated, an residual trend in Hubble residuals vs.\ $z$ might remain. If indeed $\mathcal{D}(z)$ is attributable to SNIa systematics, then our results underscore the importance of continually refining SNIa analyses \citep{LHuillier:2018rsv,DES:2024ank,Perivolaropoulos:2023iqj}. Future SNIa samples from LSST (Rubin Observatory \citep{2009arXiv0912.0201L}) and JWST \citep{2006SSRv..123..485G} can probe higher redshifts with improved calibration, allowing direct tests of whether SNIa brightness evolves. 

\textbf{Systematic Uncertainties in BAO and CMB:}  
BAO and CMB measurements are widely regarded as robust anchors of the cosmic distance scale, yet they are not immune to systematic errors. For BAO, uncertainties can arise from inaccuracies in BAO modeling which may slightly shift the measured distance observables, such as \(D_M(z)/r_d\) and \(H(z)r_d\) \citep{Ding:2017gad,Prada:2014bra,Nishimichi:2017gdq}. Additionally, the determination of the sound horizon \(r_d\) from the CMB involves assumptions about the physics of the early universe such as the baryon density and recombination history which, if are calculated incorrectly, can introduce unaccounted residual uncertainties. For CMB observations, systematic issues may stem from instrumental calibration, beam uncertainties, and the subtraction of foreground contaminants. Although these systematics are typically sub-dominant compared to the high statistical precision of modern CMB experiments, they nonetheless contribute to the overall error budget in the inferred cosmological parameters \citep{Abitbol:2015epq,Planck:2015zbi,Aumont:2018epb}.

Motivated by these considerations, we explicitly tested the impact of sound-horizon uncertainties by repeating our full analysis with an alternative value of $r_d$, as presented in Appendix~\ref{app:rd_dependence}. We find that variations in $r_d$ primarily rescale the inferred Hubble constant, in agreement with the expected relation $H_0 \propto r_d^{-1}$, while leaving the matter density largely unchanged. Owing to degeneracies between $r_d$ and the distance-duality parameters, part of this rescaling can be absorbed by $\mathcal{D}(z)$, but the qualitative behavior of the dark energy EoS parameters remains stable. In particular, once distance-duality corrections are included, the inferred $(w_0,w_a)$ consistently shift toward the $\Lambda$CDM values, demonstrating that our main conclusions are robust against plausible systematic uncertainties in the sound horizon calibration.

\textbf{Exotic Physics Affecting Distance Measures:} Alternatively, \(\mathcal{D}(z)\) may indicate new physics beyond the standard model. One possibility is a violation of the CDDR, which states that \(D_L=(1+z)^2D_A\) under photon number conservation. Physical mechanisms that could lead to a CDDR violation include photon mixing with axion-like particles or absorption by an opaque dark sector. In such scenarios, a fraction of photons could oscillate into axions over cosmological distances, resulting in an energy-independent dimming of SNIa. Although current CMB and X-ray constraints limit these effects, a percent-level reduction in flux remains plausible. In this context, a nonzero \(d_1\) may hint at a light scalar field coupling to photons, with an interaction probability that accumulates with \((1+z)\) \citep{Csaki:2001yk,Bassett:2003zw,Mirizzi:2006zy,Holanda:2012at,Mirizzi:2005ng,Hook:2021ous}. Another class of models that can produce CDDR violations involves varying speed of light (VSL) theories, such as the minimally extended VSL (meVSL) model, which modifies the standard distance-duality relation as $D_L(z) = (1+z)^2 D_A(z) [c(z)/c_0]^\alpha$ \citep{Lee:2021xwh, Lee:2020zts, Verde:2016ccp, Santos:2025gjf}. Our phenomenological parameterization could in principle capture such effects, though distinguishing VSL from astrophysical systematics would require additional high-redshift observations and consistency checks with CMB constraints.

A key difference between systematic errors and a new physical effect is that the latter should uniformly affect all luminous sources, while SNIa-specific systematics would not impact BAO or CMB observations. Since BAO and CMB data agree while SNIa do not, our results favor a scenario in which correcting for the SNIa-specific \(\mathcal{D}(z)\) restores consistency with vanilla \(\Lambda\)CDM. This outcome suggests that the discrepancies are more likely due to unaccounted astrophysical systematics rather than a universal new physics effect, although the possibility of a subtle new interaction cannot be entirely ruled out.

\section{Conclusion}
\label{sec:Conclusion}

In this work, we present a new methodology for testing the robustness of cosmological inference using diverse cosmological datasets, specifically by examining the consistency between SNIa and BAO datasets through the CDDR. Unlike the DESI analysis, which assumes perfect calibration between these datasets, our analysis introduces a redshift-dependent parameterization for possible deviations from CDDR and performs a joint inference of both cosmological and calibration parameters. This framework allows us to marginalize over unknown systematics that may otherwise bias cosmological conclusions.

\begin{table*}[ht]
\centering
\footnotesize
\setlength{\tabcolsep}{3pt}
\renewcommand{\arraystretch}{1.5}
\resizebox{\textwidth}{!}{%
  \begin{tabular}{|l|c|c|c|c|c|c|c|c|}
    \hline
    \multicolumn{7}{|c|}
    {\textbf{Constraints on cosmological and distance duality parameters for $r_d = 147.09\,\mathrm{Mpc}$}} \\
    \hline
    \shortstack{Model} &
    $H_0$ &
    $\Omega_m$ &
    $w_0$ &
    $w_a$ &
    $d_1$ &
    $d_2$ \\
    \hline
    \shortstack{$w_0$--$w_a$CDM + $(d_1,d_2)$ \\ \textit{Logarithmic parameterization}} &
    $68.794^{+0.736}_{-0.747}$ &
    $0.311^{+0.01}_{-0.01}$ &
    $-0.938^{+0.082}_{-0.082}$ &
    $-0.431^{+0.308}_{-0.329}$ &
    $0.037^{+0.144}_{-0.026}$ &
    $1.213^{+4.454}_{-1.020}$ 
    \\
    \hline
    \shortstack{$w_0$--$w_a$CDM + $(d_1,d_2)$ \\ \textit{Exponential parameterization}} &
    $68.471^{+0.782}_{-0.751}$ &
    $0.31^{+0.01}_{-0.01}$ &
    $-0.922^{+0.081}_{-0.083}$ &
    $-0.424^{+0.307}_{-0.343}$ &
    $-0.008^{+0.006}_{-0.005}$ &
    $-1.202^{+1.943}_{-1.035}$
    \\
    \hline
  \end{tabular}
}
\caption{Summary of the constraints on the Hubble constant $H_0$, matter density $\Omega_m$, dark energy EoS parameters $(w_0, w_a)$, and the distance duality coefficients $(d_1, d_2)$ for the two extended parameterizations, assuming a sound horizon scale $r_d = 149.09$ Mpc.}
\label{tab:Rd147Summary}
\end{table*}

Crucially, this paper is not intended as a critique of the DESI results, but rather as a demonstration of how cosmological inference can be systematically biased if such degeneracies between calibration and cosmological parameters are not properly accounted for. By explicitly modeling and constraining these effects, our approach offers a more robust pathway for combining heterogeneous cosmological datasets, especially in the presence of unknown astrophysical or instrumental systematics. This method is applicable broadly and sets a precedent for future analyses that aim to extract accurate and unbiased cosmological parameters from multiple independent probes.

This analysis explores the consistency test based on the CDDR of SNIa and BAO datasets used for the inference of the low redshift expansion history of the Universe by the DESI collaboration. We find that SNIa datasets and BAO datasets fails to match the fundamental CDDR and depict a statistically significant bias across most of the low redshift range. The failure of this consistency test hint towards possible unaccounted systematics present in these datasets and hence refutes the robustness of the cosmological inference obtained using DESI and SNIa in combination with the CMB information.

Our analysis points out that there exists a strong correlation between the cosmological parameters and the parameters which can capture the deviation from CDDR. As a result, presence of any effect which violates CDDR can bias the inference of cosmological parameters. On taking into account the CDDR in terms of a redshift-dependent phenomenological model to fit the data, we find that both the dark energy EoS parameters $w_0$ and $w_a$ show a strong degeneracy with the parameters capturing the CDDR deviation. Moreover, the posterior of the dark energy EoS parameters shifts towards the value of $w_0=-1$ and $w_a=0$ which is consistent with the dark energy model as cosmological constant. This implies a consistent inference of the cosmological parameters is possible when the deviations from CDDR are marginalized over.

In summary, the recent claim of evolving dark energy EoS by the DESI collaboration using the low redshift cosmological probe is subject to unaccounted systematic (astrophysical or non-astrophysical) effects. Such contamination is likely causing an inaccurate inference of the cosmological parameters from these datasets. Combining independent datasets that are impacted by systematic can cause a precise but inaccurate inference of cosmological parameters. Our study provides a clue towards this direction and in the future more elaborate study will be required to explore the reason for this effect. Also, future cosmological analysis using different independent datasets must perform CDDR consistency tests, to confirm that possible contamination from systematic effect in either datasets is not driving the cosmological results. In the future with the availability of multi-messenger cosmological datasets by the addition of gravitational wave source catalog of bright standard sirens, both accurate and precision measurement of the dark energy will be feasible \cite{Afroz:2024lou}.

\textbf{Author's note: When this study was under preparation, papers \citep{Teixeira:2025czm,Wang:2025bkk,2025arXiv250415336C} by other authors also appeared  on arXiv that indicates a similar issue with the DESI and SNIa datasets.}

\appendix
\section{Robustness to Alternative $\mathcal{D}(z)$ Parameterizations}
\label{app:Alternative_Dz}

To ensure that our main conclusions are independent of the specific functional form adopted for the distance duality coefficient $\mathcal{D}(z)$, we test two alternative parameterizations beyond the polynomial forms presented in the Section~\ref{sec:result}. We consider:

\begin{enumerate}
    \item \textbf{Exponential form:}
    \begin{equation}
    \mathcal{D}(z) = 1.0 + d_1 \, \exp(-d_2 z),
    \label{eq:Dz_exponential}
    \end{equation}
    
    \item \textbf{Logarithmic form:}
    \begin{equation}
    \mathcal{D}(z) = 1.0 + d_1 \ln(1 + d_2 z).
    \label{eq:Dz_logarithmic}
    \end{equation}
\end{enumerate}

\begin{table*}[ht]
\centering
\footnotesize
\setlength{\tabcolsep}{3pt}
\renewcommand{\arraystretch}{1.5}
\resizebox{\textwidth}{!}{%
\begin{tabular}{|l|c|c|c|c|c|c|c|}
\hline
\multicolumn{8}{|c|}{\textbf{Constraints on cosmological and distance duality parameters for $r_d = 136.4\,\mathrm{Mpc}$}} \\
\hline
Model &
$H_0$ &
$\Omega_m$ &
$w_0$ &
$w_a$ &
$d_0$ &
$d_1$ &
$d_2$ \\
\hline
\shortstack{$w_0$--$w_a$CDM + $(d_0,d_1,d_2)$ \\ \textit{Polynomial}} &
$72.23^{+0.77}_{-0.76}$ &
$0.316^{+0.010}_{-0.010}$ &
$-0.829^{+0.084}_{-0.083}$ &
$-0.536^{+0.318}_{-0.349}$ &
$1.213^{+0.121}_{-0.122}$ &
$-0.302^{+0.174}_{-0.176}$ &
$0.113^{+0.063}_{-0.065}$ \\
\hline
\shortstack{$w_0$--$w_a$CDM + $(d_1,d_2)$ \\ \textit{Logarithmic}} &
$72.289^{+0.754}_{-0.742}$ &
$0.316^{+0.010}_{-0.010}$ &
$-0.813^{+0.073}_{-0.073}$ &
$-0.511^{+0.285}_{-0.310}$ &
-- &
$0.006^{+0.068}_{-0.020}$ &
$1.090^{+4.807}_{-1.001}$ \\
\hline
\shortstack{$w_0$--$w_a$CDM + $(d_1,d_2)$ \\ \textit{Exponential}} &
$72.289^{+0.654}_{-0.615}$ &
$0.316^{+0.010}_{-0.010}$ &
$-0.821^{+0.065}_{-0.066}$ &
$-0.470^{+0.279}_{-0.317}$ &
-- &
$0.004^{+0.007}_{-0.006}$ &
$0.648^{+1.960}_{-1.838}$ \\
\hline
\end{tabular}
}
\caption{Summary of marginalized constraints on the Hubble constant $H_0$, matter density $\Omega_m$, dark energy EoS parameters $(w_0, w_a)$, and the distance duality coefficients $(d_0, d_1, d_2)$ for different $\mathcal{D}(z)$ parameterizations, assuming a sound horizon scale $r_d = 136.4\,\mathrm{Mpc}$. For the \textit{logarithmic} and \textit{exponential} parameterizations, the coefficient $d_0$ is not included and is therefore indicated by ``--''.}
\label{tab:Rd136Summary}
\end{table*}

Each parameterization provides a distinct mathematical representation of possible redshift-dependent systematics or physical effects that could lead to deviations from the standard distance duality relation. The exponential form naturally captures effects that decay with redshift (such as absorption mechanisms or evolving calibration systematics), while the logarithmic form introduces a slowly varying correction appropriate for gradual calibration drifts.

\begin{figure*}[ht]
\centering
\includegraphics[width=15cm, height=15cm]{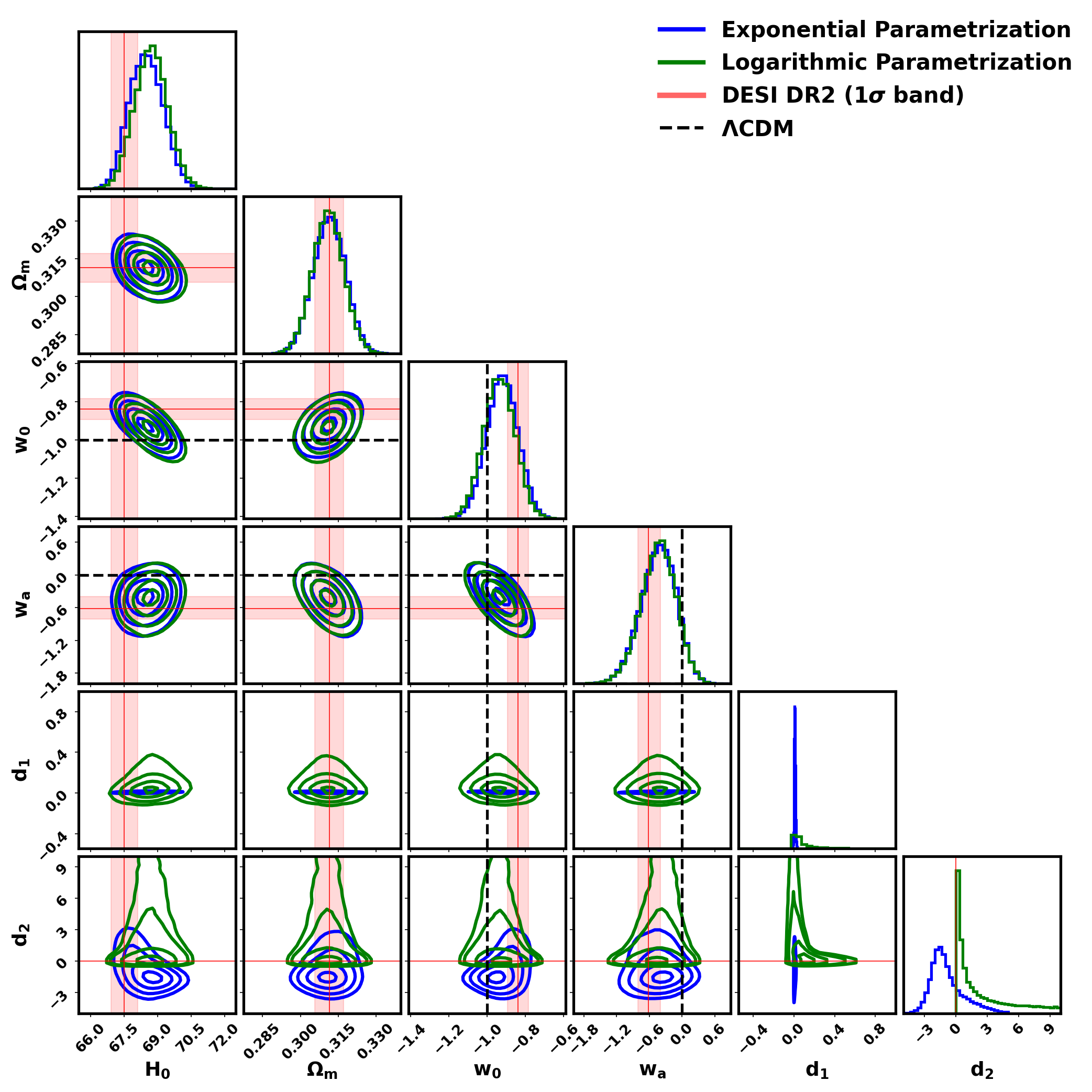}
\caption{Joint six-parameter posterior constraints on the parameter set $\{H_0,\Omega_m,w_0,w_a,d_1,d_2\}$ in a flat $w_0$-$w_a$CDM model including distance-duality corrections for the \textit{logarithmic} and \textit{exponential} forms, assuming $r_d = 147.09\,\mathrm{Mpc}$. The blue contours show the joint constraints from DESI DR2 + Pantheon+ for the extended \textit{exponential} parameterization, while the green contours correspond to the extended \textit{logarithmic} parameterization. The red shaded regions represent the DESI DR2 (BAO + SNIa + Planck $\Omega_m$ prior) constraints. We impose a Gaussian Planck prior on $\Omega_m$ (mean $=0.315$, $\sigma = 0.007$), flat priors on $H_0 \in [60,80]$, $w_0 \in [-2,1]$, and $w_a \in [-3,3]$ with the requirement $w_0 + w_a < 0$, and flat priors on the distance-duality coefficients $d_1 \in [-1.0,1.0]$ and $d_2 \in [0,10]$ for the \textit{logarithmic} form and $d_2 \in [-5,5]$ for the \textit{exponential} form.}

\label{fig:ExponentialCase}
\end{figure*}

For each functional form, we perform a full Bayesian analysis as described in Section~\ref{sec:method}, using the combined DESI DR2 + Pantheon+ dataset and adopting the same priors on the cosmological parameters. Specifically, we impose a Gaussian prior on $\Omega_m$ with mean $=0.315$ and standard deviation $ = 0.007$, and flat priors on $H_0 \in [60,80]$, $w_0 \in [-2,1]$, and $w_a \in [-3,3]$, with the additional constraint $w_0 + w_a < 0$. For the $\mathcal{D}(z)$ parameters, we adopt flat priors with $d_1 \in [-1.0,1.0]$, while $d_2 \in [0,10]$ for the \textit{logarithmic} parameterization and $d_2 \in [-5.0,5.0]$ for the \textit{exponential} parameterization. Table~\ref{tab:Rd147Summary} summarizes the constraints on the cosmological and distance duality parameters for both parameterizations, and Figure~\ref{fig:ExponentialCase} shows the full posterior distributions of the six parameters.

The results demonstrate remarkable consistency across different functional forms. In both the \textit{exponential} and \textit{logarithmic} cases, the inclusion of $\mathcal{D}(z)$ corrections systematically shifts the dark energy parameters $(w_0, w_a)$ closer to their $\Lambda$CDM values $(-1, 0)$ compared to the baseline analysis, independent of the specific functional form adopted. Together with the polynomial parameterization $\mathcal{D}(z) = d_0 + d_1(1+z) + d_2(1+z)^2$ presented in the main text, these results confirm that alternative functional choices lead to both qualitatively and quantitatively consistent conclusions, thereby reinforcing the robustness of our main scientific results.

\section{Dependence on the Sound Horizon Scale $r_d$}
\label{app:rd_dependence}

To examine the sensitivity of our cosmological inferences to the assumed sound horizon scale, we repeat the full analysis using an alternative value of the sound horizon, $r_d = 136.4\,\mathrm{Mpc}$.

\begin{figure*}[ht]
\centering
\includegraphics[width=14cm, height=14cm]{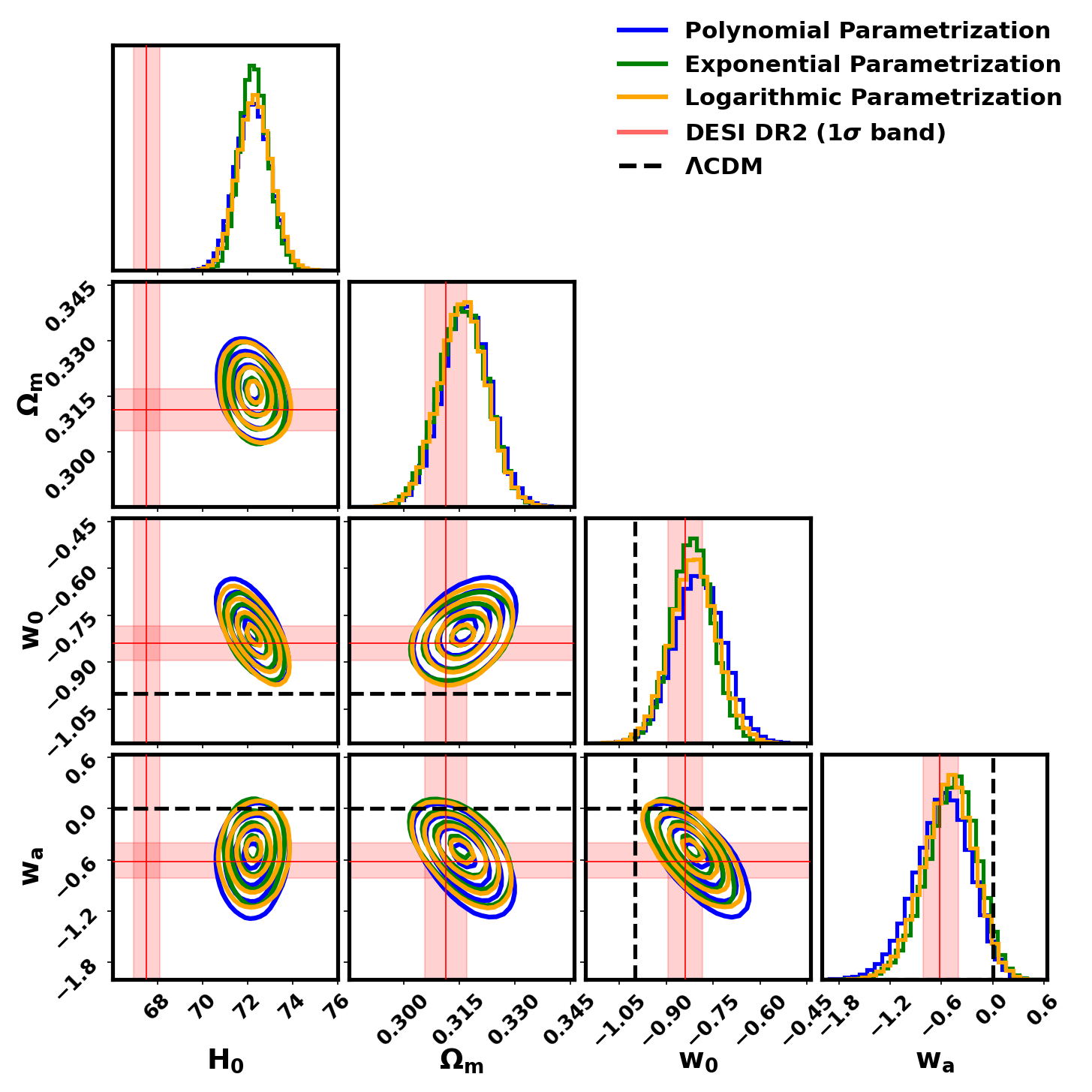}
\caption{Marginalized posterior distributions of the cosmological parameters $\{H_0, \Omega_m, w_0, w_a\}$ obtained assuming an alternative sound horizon scale $r_d = 136.4\,\mathrm{Mpc}$. The contours correspond to: (i) a joint seven-parameter analysis $\{H_0, \Omega_m, w_0, w_a, d_0, d_1, d_2\}$ using the \textit{polynomial} distance duality parameterization, and (ii) joint six-parameter analyses $\{H_0, \Omega_m, w_0, w_a, d_1, d_2\}$ using the \textit{logarithmic} and \textit{exponential} parameterizations. All analyses combine DESI DR2 BAO and Pantheon+ SNIa data, assuming a flat $w_0$--$w_a$CDM cosmology. A Gaussian Planck prior on the matter density is imposed, $\Omega_m = 0.315 \pm 0.007$, together with flat priors $H_0 \in [60,80]$, $w_0 \in [-2,1]$, and $w_a \in [-3,3]$, subject to the condition $w_0 + w_a < 0$. Flat priors are adopted for the distance duality parameters: $d_0 \in [-1,2]$ (polynomial only), $d_1 \in [-1,1]$ (all parameterizations), $d_2 \in [-5,5]$ (polynomial and exponential), and $d_2 \in [0,10]$ (logarithmic). The dashed lines in the diagonal panels indicate the $\Lambda$CDM values $w_0 = -1$ and $w_a = 0$, as well as the standard CDDR expectation.}
\label{fig:rd_alternative}
\end{figure*}

We perform a joint seven-parameter analysis $\{H_0, \Omega_m, w_0, w_a, d_0, d_1, d_2\}$ using the \textit{polynomial} parameterization of the distance duality coefficient $\mathcal{D}(z)$. In addition, we carry out joint six-parameter analyses $\{H_0, \Omega_m, w_0, w_a, d_1, d_2\}$ employing the \textit{logarithmic} and \textit{exponential} parameterizations. All analyses are carried out within a spatially flat $w_0$--$w_a$CDM framework. The adopted priors for all cosmological and distance duality parameters are summarized in the caption of Fig.~\ref{fig:rd_alternative}.

For clarity of presentation, Fig.~\ref{fig:rd_alternative} displays only the marginalized posterior distributions of the four cosmological parameters  $\{H_0, \Omega_m, w_0, w_a\}$, while the inference is performed jointly over the full parameter space, including all distance-duality parameters, whose marginalized constraints are summarized in Table~\ref{tab:Rd136Summary}.

As expected from the inverse scaling relation $H_0 \propto r_d^{-1}$, lowering the sound horizon leads to a systematic shift of the inferred Hubble constant toward higher values. The dark energy EoS parameters $(w_0, w_a)$ also exhibit mild shifts towards higher value than with the $r_d=147.09$ Mpc; however, the trend of posterior shifts towards the $\Lambda$CDM values for $(w_0, w_a)$ is consistent. These changes arise from correlations between $H_0$, the distance duality parameters, and the dark energy sector, as well as from the imposed Gaussian Planck prior on $\Omega_m$.

Physically, this robustness can be understood from the fact that the sound horizon primarily sets the overall cosmological distance scale. As a result, variations in $r_d$ are largely absorbed by the Hubble constant $H_0$, while the distance duality parameters and dark energy EoS parameters contribute only subdominantly. Consequently, the impact of changing $r_d$ is significantly larger for $H_0$ than for the other cosmological parameters.

\appendix

\section*{Acknowledgements}
We thank the referee for constructive and insightful comments, in particular for suggesting the exploration of additional parameterizations of the cosmic distance duality relation and the assessment of the impact of alternative choices of the sound horizon scale, which significantly improved the robustness and clarity of this work. This work is part of the \texttt{⟨data|theory⟩ Universe-Lab}, supported by TIFR and the Department of Atomic Energy, Government of India. The authors thank the computing cluster of the \texttt{⟨data|theory⟩ Universe-Lab} for providing computing resources. We acknowledge the use of publicly available data from the Pantheon\(+\) supernova sample \citep{Brout:2022vxf} and the DESI BAO measurements \citep{DESI:2025zgx}, as well as several software packages that significantly facilitated this analysis: Astropy \citep{price2018astropy}, Pandas \citep{mckinney2011pandas}, NumPy \citep{harris2020array}, Seaborn \citep{bisong2019matplotlib}, SciPy \citep{virtanen2020scipy}, emcee \citep{foreman2013emcee}, pygtc \citep{bocquet2019pygtc}, corner \citep{corner}, and Matplotlib \citep{Hunter:2007}.

\bibliographystyle{unsrt}
\bibliography{references}

\begin{thebibliography}{10}

\bibitem{2013PhR...530...87W}
David~H. {Weinberg}, Michael~J. {Mortonson}, Daniel~J. {Eisenstein}, Christopher {Hirata}, Adam~G. {Riess}, and Eduardo {Rozo}.
\newblock {Observational probes of cosmic acceleration}.
\newblock {\em \physrep}, 530(2):87--255, September 2013.

\bibitem{Lahav:2024npe}
Ofer Lahav and Andrew~R. Liddle.
\newblock {The Cosmological Parameters (2023)}.
\newblock {\em arXiv e-prints}, 3 2024.

\bibitem{Staicova:2025huq}
Denitsa Staicova.
\newblock {Modern Bayesian Sampling Methods for Cosmological Inference: A Comparative Study}.
\newblock {\em Universe}, 11(2):68, 2025.

\bibitem{LuisBernal:2018drn}
Jos\'e Luis~Bernal and John~A. Peacock.
\newblock {Conservative cosmology: combining data with allowance for unknown systematics}.
\newblock {\em JCAP}, 07:002, 2018.

\bibitem{Piras:2024dml}
Davide Piras, Alicja Polanska, Alessio Spurio~Mancini, Matthew~A. Price, and Jason~D. McEwen.
\newblock {The future of cosmological likelihood-based inference: accelerated high-dimensional parameter estimation and model comparison}.
\newblock {\em arXiv e-prints}, 5 2024.

\bibitem{Steinhardt:2025znn}
Charles~L. Steinhardt, Preston Phillips, and Radoslaw Wojtak.
\newblock {Dark Energy Constraints and Joint Cosmological Inference from Mutually Inconsistent Observations}.
\newblock {\em arXiv e-prints}, 4 2025.

\bibitem{Abdalla:2022yfr}
Elcio Abdalla et~al.
\newblock {Cosmology intertwined: A review of the particle physics, astrophysics, and cosmology associated with the cosmological tensions and anomalies}.
\newblock {\em JHEAp}, 34:49--211, 2022.

\bibitem{Carr:2021lcj}
Anthony Carr, Tamara~M. Davis, Dan Scolnic, Daniel Scolnic, Khaled Said, Dillon Brout, Erik~R. Peterson, and Richard Kessler.
\newblock {The Pantheon+ analysis: Improving the redshifts and peculiar velocities of Type Ia supernovae used in cosmological analyses}.
\newblock {\em Publ. Astron. Soc. Austral.}, 39:e046, 2022.

\bibitem{DES:2018paw}
T.~M.~C. Abbott et~al.
\newblock {First Cosmology Results using Type Ia Supernovae from the Dark Energy Survey: Constraints on Cosmological Parameters}.
\newblock {\em Astrophys. J. Lett.}, 872(2):L30, 2019.

\bibitem{DESI:2025zgx}
M.~Abdul~Karim et~al.
\newblock {DESI DR2 Results II: Measurements of Baryon Acoustic Oscillations and Cosmological Constraints}.
\newblock {\em arXiv e-prints}, 3 2025.

\bibitem{Planck:2018vyg}
N.~Aghanim et~al.
\newblock {Planck 2018 results. VI. Cosmological parameters}.
\newblock {\em Astron. Astrophys.}, 641:A6, 2020.
\newblock [Erratum: Astron.Astrophys. 652, C4 (2021)].

\bibitem{Lemos:2023rdh}
Pablo Lemos and Paul Shah.
\newblock {The Cosmic Microwave Background and $H_0$}.
\newblock {\em arXiv e-prints}, 7 2023.

\bibitem{ACT:2025fju}
Thibaut Louis et~al.
\newblock {The Atacama Cosmology Telescope: DR6 Power Spectra, Likelihoods and $\Lambda$CDM Parameters}.
\newblock {\em arXiv e-prints}, 3 2025.

\bibitem{DiValentino:2025sru}
Eleonora Di~Valentino et~al.
\newblock {The CosmoVerse White Paper: Addressing observational tensions in cosmology with systematics and fundamental physics}.
\newblock {\em arXiv e-prints}, 4 2025.

\bibitem{DES:2018gui}
T.~M.~C. Abbott et~al.
\newblock {The Dark Energy Survey Data Release 1}.
\newblock {\em Astrophys. J. Suppl.}, 239(2):18, 2018.

\bibitem{DESI:2016fyo}
Amir Aghamousa et~al.
\newblock {The DESI Experiment Part I: Science,Targeting, and Survey Design}.
\newblock {\em arXiv e-prints}, 10 2016.

\bibitem{Brout:2022vxf}
Dillon Brout et~al.
\newblock {The Pantheon+ Analysis: Cosmological Constraints}.
\newblock {\em Astrophys. J.}, 938(2):110, 2022.

\bibitem{Rubin:2023ovl}
David Rubin et~al.
\newblock {Union Through UNITY: Cosmology with 2,000 SNe Using a Unified Bayesian Framework}.
\newblock {\em arXiv e-prints}, 11 2023.

\bibitem{DES:2024jxu}
T.~M.~C. Abbott et~al.
\newblock {The Dark Energy Survey: Cosmology Results with \ensuremath{\sim}1500 New High-redshift Type Ia Supernovae Using the Full 5 yr Data Set}.
\newblock {\em Astrophys. J. Lett.}, 973(1):L14, 2024.

\bibitem{Chevallier:2000qy}
Michel Chevallier and David Polarski.
\newblock {Accelerating universes with scaling dark matter}.
\newblock {\em Int. J. Mod. Phys. D}, 10:213--224, 2001.

\bibitem{Linder:2002et}
Eric~V. Linder.
\newblock {Exploring the expansion history of the universe}.
\newblock {\em Phys. Rev. Lett.}, 90:091301, 2003.

\bibitem{dePutter:2008wt}
Roland de~Putter and Eric~V. Linder.
\newblock {Calibrating Dark Energy}.
\newblock {\em JCAP}, 10:042, 2008.

\bibitem{DESI:2024uvr}
A.~G. Adame et~al.
\newblock {DESI 2024 III: baryon acoustic oscillations from galaxies and quasars}.
\newblock {\em JCAP}, 04:012, 2025.

\bibitem{Shlivko:2024llw}
David Shlivko and Paul~J. Steinhardt.
\newblock {Assessing observational constraints on dark energy}.
\newblock {\em Phys. Lett. B}, 855:138826, 2024.

\bibitem{PhysRevD.108.103519}
William~J. Wolf and Pedro~G. Ferreira.
\newblock Underdetermination of dark energy.
\newblock {\em Phys. Rev. D}, 108:103519, Nov 2023.

\bibitem{Afroz:2024lou}
Samsuzzaman Afroz and Suvodip Mukherjee.
\newblock {Multi-messenger cosmology: A route to accurate inference of dark energy beyond CPL parametrization from XG detectors}.
\newblock {\em JCAP}, 03:070, 2025.

\bibitem{Colgain:2024xqj}
Eoin~\'O. Colg\'ain, Maria~Giovanna Dainotti, Salvatore Capozziello, Saeed Pourojaghi, M.~M. Sheikh-Jabbari, and Dejan Stojkovic.
\newblock {Does DESI 2024 Confirm $\Lambda$CDM?}
\newblock {\em arXiv e-prints}, 4 2024.

\bibitem{Colgain:2024ksa}
Eoin~\'O. Colg\'ain, Saeed Pourojaghi, and M.~M. Sheikh-Jabbari.
\newblock {Implications of DES 5YR SNe Dataset for $\Lambda $CDM}.
\newblock {\em Eur. Phys. J. C}, 85(3):286, 2025.

\bibitem{Colgain:2024mtg}
Eoin~\'O. Colg\'ain and M.~M. Sheikh-Jabbari.
\newblock {DESI and SNe: Dynamical Dark Energy, $\Omega_m$ Tension or Systematics?}
\newblock {\em arXiv e-prints}, 12 2024.

\bibitem{Colgain:2025nzf}
Eoin~\'O. Colg\'ain, Saeed Pourojaghi, M.~M. Sheikh-Jabbari, and Lu~Yin.
\newblock {How much has DESI dark energy evolved since DR1?}
\newblock {\em arXiv e-prints}, 4 2025.

\bibitem{Holanda:2010vb}
R.~F.~L. Holanda, J.~A.~S. Lima, and M.~B. Ribeiro.
\newblock {Testing the Distance-Duality Relation with Galaxy Clusters and Type Ia Supernovae}.
\newblock {\em Astrophys. J. Lett.}, 722:L233--L237, 2010.

\bibitem{Liao:2015uzb}
Kai Liao, Zhengxiang Li, Shuo Cao, Marek Biesiada, Xiaogang Zheng, and Zong-Hong Zhu.
\newblock {The Distance Duality Relation From Strong Gravitational Lensing}.
\newblock {\em Astrophys. J.}, 822(2):74, 2016.

\bibitem{Keil:2025ysb}
Felicitas Keil, Savvas Nesseris, Isaac Tutusaus, and Alain Blanchard.
\newblock {Probing the Distance Duality Relation with Machine Learning and Recent Data}.
\newblock {\em arXiv e-prints}, 4 2025.

\bibitem{2007GReGr..39.1055E}
I.~M.~H. {Etherington}.
\newblock {Republication of: LX. On the definition of distance in general relativity}.
\newblock {\em General Relativity and Gravitation}, 39(7):1055--1067, July 2007.

\bibitem{Afroz:2024joi}
Samsuzzaman Afroz and Suvodip Mukherjee.
\newblock {Prospect of precision cosmology and testing general relativity using binary black holes \textendash{} galaxies cross-correlation}.
\newblock {\em Mon. Not. Roy. Astron. Soc.}, 534(2):1283--1298, 2024.

\bibitem{Afroz:2024oui}
Samsuzzaman Afroz and Suvodip Mukherjee.
\newblock {A model-independent precision test of General Relativity using LISA bright standard sirens}.
\newblock {\em JCAP}, 10:100, 2024.

\bibitem{Afroz:2023ndy}
Samsuzzaman Afroz and Suvodip Mukherjee.
\newblock {A model-independent precision test of general relativity using bright standard sirens from ongoing and upcoming detectors}.
\newblock {\em Mon. Not. Roy. Astron. Soc.}, 530(4):3812--3826, 2024.

\bibitem{Mukherjee:2020mha}
Suvodip Mukherjee, Benjamin~D. Wandelt, and Joseph Silk.
\newblock {Testing the general theory of relativity using gravitational wave propagation from dark standard sirens}.
\newblock {\em Mon. Not. Roy. Astron. Soc.}, 502(1):1136--1144, 2021.

\bibitem{2019asclsoft10019T}
Jes{\'u}s {Torrado} and Antony {Lewis}.
\newblock {Cobaya: Bayesian analysis in cosmology}.
\newblock Astrophysics Source Code Library, record ascl:1910.019, October 2019.

\bibitem{Lewis:1999bs}
Antony Lewis, Anthony Challinor, and Anthony Lasenby.
\newblock {Efficient computation of CMB anisotropies in closed FRW models}.
\newblock {\em Astrophys. J.}, 538:473--476, 2000.

\bibitem{dickey1971weighted}
James~M Dickey.
\newblock The weighted likelihood ratio, linear hypotheses on normal location parameters.
\newblock {\em The Annals of Mathematical Statistics}, pages 204--223, 1971.

\bibitem{Trotta:2008qt}
Roberto Trotta.
\newblock {Bayes in the sky: Bayesian inference and model selection in cosmology}.
\newblock {\em Contemp. Phys.}, 49:71--104, 2008.

\bibitem{2010ApJ...711L..66B}
E.~{Bravo}, I.~{Dom{\'\i}nguez}, C.~{Badenes}, L.~{Piersanti}, and O.~{Straniero}.
\newblock {Metallicity as a Source of Dispersion in the SNIa Bolometric Light Curve Luminosity-Width Relationship}.
\newblock {\em \apjl}, 711(2):L66--L70, March 2010.

\bibitem{2011MNRAS.414.1592B}
E.~{Bravo} and C.~{Badenes}.
\newblock {Is the metallicity of their host galaxies a good measure of the metallicity of Type Ia supernovae?}
\newblock {\em \mnras}, 414(2):1592--1606, June 2011.

\bibitem{Riess:2006yk}
Adam~G Riess and Mario Livio.
\newblock {The first type ia supernovae: an empirical approach to taming evolutionary effects in dark energy surveys from sne ia at z\ensuremath{>}2}.
\newblock {\em Astrophys. J.}, 648:884--889, 2006.

\bibitem{Panagia:2007eg}
Nino Panagia, Massimo Della~Valle, and Filippo Mannucci.
\newblock {Type Ia Supernova Rates Near and Far}.
\newblock {\em AIP Conf. Proc.}, 924(1):373--382, 2007.

\bibitem{Jones:2013dta}
David~O. Jones et~al.
\newblock {The Discovery of the Most Distant Known Type Ia Supernova at Redshift 1.914}.
\newblock {\em Astrophys. J.}, 768:166, 2013.

\bibitem{2007A&A...464..465R}
A.~R. {Robaina} and J.~{Cepa}.
\newblock {Redshift-distance relations from type Ia supernova observations. New constraints on grey dust models}.
\newblock {\em \aap}, 464(2):465--470, March 2007.

\bibitem{Goobar:2018smm}
Ariel Goobar, Suhail Dhawan, and Daniel Scolnic.
\newblock {The cosmic transparency measured with Type Ia supernovae: implications for intergalactic dust}.
\newblock {\em Mon. Not. Roy. Astron. Soc.}, 477(1):L75--L79, 2018.

\bibitem{Vavrycuk:2019czf}
Vaclav Vavrycuk.
\newblock {Universe opacity and Type Ia supernova dimming}.
\newblock {\em Mon. Not. Roy. Astron. Soc.}, 489(1):L63--L68, 2019.

\bibitem{Ostman:2004eh}
Linda Ostman and Edvard Mortsell.
\newblock {Limiting the dimming of distant Type Ia supernovae}.
\newblock {\em JCAP}, 02:005, 2005.

\bibitem{Croft:2000ns}
Rupert A.~C. Croft, Romeel Dave, Lars Hernquist, and Neal Katz.
\newblock {Simulating the effects of intergalactic grey dust}.
\newblock {\em Astrophys. J. Lett.}, 534:L123, 2000.

\bibitem{Brout:2021mpj}
Dillon Brout et~al.
\newblock {The Pantheon+ Analysis: SuperCal-fragilistic Cross Calibration, Retrained SALT2 Light-curve Model, and Calibration Systematic Uncertainty}.
\newblock {\em Astrophys. J.}, 938(2):111, 2022.

\bibitem{Brownsberger:2021uue}
Sasha~R. Brownsberger, Dillon Brout, Daniel Scolnic, Christopher~W. Stubbs, and Adam~G. Riess.
\newblock {Dependence of Cosmological Constraints on Gray Photometric Zero-point Uncertainties of Supernova Surveys}.
\newblock {\em Astrophys. J.}, 944(2):188, 2023.

\bibitem{Riess:2024vfa}
Adam~G. Riess et~al.
\newblock {JWST Validates HST Distance Measurements: Selection of Supernova Subsample Explains Differences in JWST Estimates of Local H $_{0}$}.
\newblock {\em Astrophys. J.}, 977(1):120, 2024.

\bibitem{LHuillier:2018rsv}
Benjamin L'Huillier, Arman Shafieloo, Eric~V. Linder, and Alex~G. Kim.
\newblock {Model Independent Expansion History from Supernovae: Cosmology versus Systematics}.
\newblock {\em Mon. Not. Roy. Astron. Soc.}, 485(2):2783--2790, 2019.

\bibitem{DES:2024ank}
M.~Toy et~al.
\newblock {Reduction of the type Ia supernova host galaxy step in the outer regions of galaxies}.
\newblock {\em Mon. Not. Roy. Astron. Soc.}, 538(1):181--197, 2025.

\bibitem{Perivolaropoulos:2023iqj}
Leandros Perivolaropoulos and Foteini Skara.
\newblock {On the homogeneity of SnIa absolute magnitude in the Pantheon+~sample}.
\newblock {\em Mon. Not. Roy. Astron. Soc.}, 520(4):5110--5125, 2023.

\bibitem{2009arXiv0912.0201L}
{LSST Science Collaboration}.
\newblock {LSST Science Book, Version 2.0}.
\newblock {\em arXiv e-prints}, page arXiv:0912.0201, December 2009.

\bibitem{2006SSRv..123..485G}
Jonathan~P. {Gardner}, John~C. {Mather}, Mark {Clampin}, Rene {Doyon}, Matthew~A. {Greenhouse}, Heidi~B. {Hammel}, John~B. {Hutchings}, Peter {Jakobsen}, Simon~J. {Lilly}, Knox~S. {Long}, Jonathan~I. {Lunine}, Mark~J. {McCaughrean}, Matt {Mountain}, John {Nella}, George~H. {Rieke}, Marcia~J. {Rieke}, Hans-Walter {Rix}, Eric~P. {Smith}, George {Sonneborn}, Massimo {Stiavelli}, H.~S. {Stockman}, Rogier~A. {Windhorst}, and Gillian~S. {Wright}.
\newblock {The James Webb Space Telescope}.
\newblock {\em \ssr}, 123(4):485--606, April 2006.

\bibitem{Ding:2017gad}
Zhejie Ding, Hee-Jong Seo, Zvonimir Vlah, Yu~Feng, Marcel Schmittfull, and Florian Beutler.
\newblock {Theoretical Systematics of Future Baryon Acoustic Oscillation Surveys}.
\newblock {\em Mon. Not. Roy. Astron. Soc.}, 479(1):1021--1054, 2018.

\bibitem{Prada:2014bra}
Francisco Prada, Claudia~G. Sc\'occola, Chia-Hsun Chuang, Gustavo Yepes, Anatoly~A. Klypin, Francisco-Shu Kitaura, Stefan Gottl\"ober, and Cheng Zhao.
\newblock {Hunting down systematics in baryon acoustic oscillations after cosmic high noon}.
\newblock {\em Mon. Not. Roy. Astron. Soc.}, 458(1):613--623, 2016.

\bibitem{Nishimichi:2017gdq}
Takahiro Nishimichi, Eugenio Noda, Marco Peloso, and Massimo Pietroni.
\newblock {BAO Extractor: bias and redshift space effects}.
\newblock {\em JCAP}, 01:035, 2018.

\bibitem{Abitbol:2015epq}
Maximilian~H. Abitbol, J.~C. Hill, and B.~R. Johnson.
\newblock {Foreground-Induced Biases in CMB Polarimeter Self-Calibration}.
\newblock {\em Mon. Not. Roy. Astron. Soc.}, 457(2):1796--1803, 2016.

\bibitem{Planck:2015zbi}
P.~A.~R. Ade et~al.
\newblock {Planck 2015 results. III. LFI systematic uncertainties}.
\newblock {\em Astron. Astrophys.}, 594:A3, 2016.

\bibitem{Aumont:2018epb}
Jonathan Aumont, Juan~Francisco Mac\'\i{}as-P\'erez, Alessia Ritacco, Nicolas Ponthieu, and Anna Mangilli.
\newblock {Absolute calibration of the polarisation angle for future CMB $B$-mode experiments from current and future measurements of the Crab nebula}.
\newblock {\em Astron. Astrophys.}, 634:A100, 2020.

\bibitem{Csaki:2001yk}
Csaba Csaki, Nemanja Kaloper, and John Terning.
\newblock {Dimming supernovae without cosmic acceleration}.
\newblock {\em Phys. Rev. Lett.}, 88:161302, 2002.

\bibitem{Bassett:2003zw}
Bruce~A. Bassett.
\newblock {Cosmic acceleration vs axion - photon mixing}.
\newblock {\em Astrophys. J.}, 607:661--664, 2004.

\bibitem{Mirizzi:2006zy}
Alessandro Mirizzi, Georg~G. Raffelt, and Pasquale~D. Serpico.
\newblock {Photon-axion conversion in intergalactic magnetic fields and cosmological consequences}.
\newblock {\em Lect. Notes Phys.}, 741:115--134, 2008.

\bibitem{Holanda:2012at}
R.~F.~L. Holanda, R.~S. Gon\c{c}alves, and J.~S. Alcaniz.
\newblock {A test for cosmic distance duality}.
\newblock {\em JCAP}, 06:022, 2012.

\bibitem{Mirizzi:2005ng}
Alessandro Mirizzi, Georg~G. Raffelt, and Pasquale~D. Serpico.
\newblock {Photon-axion conversion as a mechanism for supernova dimming: Limits from CMB spectral distortion}.
\newblock {\em Phys. Rev. D}, 72:023501, 2005.

\bibitem{Hook:2021ous}
Anson Hook, Gustavo Marques-Tavares, and Clayton Ristow.
\newblock {Supernova constraints on an axion-photon-dark photon interaction}.
\newblock {\em JHEP}, 06:167, 2021.

\bibitem{Lee:2021xwh}
Seokcheon Lee.
\newblock {Cosmic distance duality as a probe of minimally extended varying speed of light}.
\newblock 8 2021.

\bibitem{Lee:2020zts}
Seokcheon Lee.
\newblock {The minimally extended Varying Speed of Light (meVSL)}.
\newblock {\em JCAP}, 08:054, 2021.

\bibitem{Verde:2016ccp}
Licia Verde, Jose~Luis Bernal, Alan~F. Heavens, and Raul Jimenez.
\newblock {The length of the low-redshift standard ruler}.
\newblock {\em Mon. Not. Roy. Astron. Soc.}, 467(1):731--736, 2017.

\bibitem{Santos:2025gjf}
Jaiane Santos, Carlos Bengaly, and Rodrigo~S. Gon{\c{c}}alves.
\newblock {Current constraints on the minimally extended varying speed of light model through the cosmic distance duality relation}.
\newblock {\em JCAP}, 11:086, 2025.

\bibitem{Teixeira:2025czm}
Elsa~M. Teixeira, William Giar\`e, Natalie~B. Hogg, Thomas Montandon, Ad\`ele Poudou, and Vivian Poulin.
\newblock {Implications of distance duality violation for the $H_0$ tension and evolving dark energy}.
\newblock {\em arXiv e-prints}, 4 2025.

\bibitem{Wang:2025bkk}
Deng Wang and David Mota.
\newblock {Did DESI DR2 truly reveal dynamical dark energy?}
\newblock {\em arXiv e-prints}, 4 2025.

\bibitem{2025arXiv250415336C}
Marina {Cort{\^e}s} and Andrew~R {Liddle}.
\newblock {On DESI's DR2 exclusion of $\Lambda$CDM}.
\newblock {\em arXiv e-prints}, page arXiv:2504.15336, April 2025.

\bibitem{price2018astropy}
Adrian~M Price-Whelan, BM~Sip{\H{o}}cz, HM~G{\"u}nther, PL~Lim, SM~Crawford, S~Conseil, DL~Shupe, MW~Craig, N~Dencheva, A~Ginsburg, et~al.
\newblock The astropy project: building an open-science project and status of the v2. 0 core package.
\newblock {\em The Astronomical Journal}, 156(3):123, 2018.

\bibitem{mckinney2011pandas}
Wes McKinney et~al.
\newblock pandas: a foundational python library for data analysis and statistics.
\newblock {\em Python for high performance and scientific computing}, 14(9):1--9, 2011.

\bibitem{harris2020array}
Charles~R Harris, K~Jarrod Millman, St{\'e}fan~J Van Der~Walt, Ralf Gommers, Pauli Virtanen, David Cournapeau, Eric Wieser, Julian Taylor, Sebastian Berg, Nathaniel~J Smith, et~al.
\newblock Array programming with numpy.
\newblock {\em Nature}, 585(7825):357--362, 2020.

\bibitem{bisong2019matplotlib}
Ekaba Bisong and Ekaba Bisong.
\newblock Matplotlib and seaborn.
\newblock {\em Building Machine Learning and Deep Learning Models on Google Cloud Platform: A Comprehensive Guide for Beginners}, pages 151--165, 2019.

\bibitem{virtanen2020scipy}
Pauli Virtanen, Ralf Gommers, Travis~E Oliphant, Evgeni Burovski, David Cournapeau, Warren Weckesser, Pearu Peterson, Stefan van~der Walt, Denis Laxalde, Matthew Brett, et~al.
\newblock Scipy/scipy: Scipy 0.19. 0.
\newblock {\em Zenodo}, 2020.

\bibitem{foreman2013emcee}
Daniel Foreman-Mackey, David~W Hogg, Dustin Lang, and Jonathan Goodman.
\newblock emcee: the mcmc hammer.
\newblock {\em Publications of the Astronomical Society of the Pacific}, 125(925):306, 2013.

\bibitem{bocquet2019pygtc}
Sebastian Bocquet and Faustin~W Carter.
\newblock pygtc: Parameter covariance plots.
\newblock {\em Astrophysics Source Code Library}, pages ascl--1907, 2019.

\bibitem{corner}
Daniel Foreman-Mackey.
\newblock corner.py: Scatterplot matrices in python.
\newblock {\em The Journal of Open Source Software}, 1(2):24, jun 2016.

\bibitem{Hunter:2007}
J.~D. Hunter.
\newblock Matplotlib: A 2d graphics environment.
\newblock {\em Computing in Science \& Engineering}, 9(3):90--95, 2007.

\end{thebibliography}
\end{document}